\documentclass[aps,prx,superscriptaddress,amsmath,amssymb,twocolumn,10pt, epsfig]{revtex4-1}

\usepackage{appendix}
\usepackage{amssymb, amsmath,amsthm}    
\usepackage{graphicx}   
\usepackage{verbatim}   
\usepackage{color}      
\usepackage{subcaption} 
\usepackage{bbm}        
\usepackage{caption}
\usepackage{tikz}
\usepackage{pgflibraryarrows}
\usepackage{pgflibrarysnakes}
\usepackage{lipsum}
\usepackage{mathtools}
\usepackage{ae} 
\usepackage[T1]{fontenc}
\usepackage[pdfstartview=FitV,linktocpage,colorlinks=true,linkcolor=darkblue,citecolor=darkblue,urlcolor=darkblue]{hyperref}
\usepackage{epsfig}
\usepackage{enumerate}
\usepackage{varioref}
\usepackage{cleveref}
\usepackage{makeidx}
\makeindex
\usepackage[english]{babel}
\usepackage[normalem]{ulem}
\usepackage{physics}
\usepackage{bbold}

\newcommand{\beq}{\begin{equation}}
\newcommand{\eeq}{\end{equation}}
\newcommand{\beqq}{\begin{equation*}}
\newcommand{\eeqq}{\end{equation*}}
\newcommand\bea{\begin{array}}
\newcommand\eea{\end{array}}
\newcommand\beaa{\begin{array*}}
\newcommand\eeaa{\end{array*}}
\newcommand\beal{\begin{align}}
\newcommand\eeal{\end{align}}
\newcommand\beall{\begin{align*}}
\newcommand\eeall{\end{align*}}

\def\o{{\omega}}
\def\a{{\alpha}}

\def\s{\sigma}
\def\th{\theta}

\def\g{\gamma}
\newcommand{\ep}{\epsilon}
\newcommand{\eps}{\varepsilon}

\def\O{{\mathcal O}}

\newcommand{\td}{\tilde}

\DeclareMathOperator{\sign}{sgn}

\def\[{\left[}
\def\]{\right]}
\def\({\left(}
\def\){\right)}
\def\<{\langle}
\def\>{\rangle}

\newcommand{\nn}{\nonumber}

\definecolor{darkblue}{cmyk}{0.9,0.9,0,0}
\definecolor{greennote}{RGB}{0,135,41}

\begin{document}
	\title{The Hubbard model on the Bethe lattice via variational uniform tree states: metal-insulator transition and a Fermi liquid}
	\author{Peter Lunts}
	\email{plunts@flatironinstitute.org}
	\affiliation{Center for Computational Quantum Physics, Flatiron Institute, 162 5th Avenue, New York, NY 10010, USA}
	\author{Antoine Georges}
	\affiliation{Center for Computational Quantum Physics, Flatiron Institute, 162 5th Avenue, New York, NY 10010, USA}
\affiliation{Coll{\`e}ge de France, 11 place Marcelin Berthelot, 75005 Paris, France}
\affiliation{CPHT, CNRS, {\'E}cole Polytechnique, IP Paris, F-91128 Palaiseau, France}
\affiliation{DQMP, Universit{\'e} de Gen{\`e}ve, 24 quai Ernest Ansermet, CH-1211 Gen{\`e}ve, Suisse}
	\email{plunts@flatironinstitute.org}
	\author{E. Miles Stoudenmire}
	\affiliation{Center for Computational Quantum Physics, Flatiron Institute, 162 5th Avenue, New York, NY 10010, USA}
	\author{Matthew Fishman}
	\affiliation{Center for Computational Quantum Physics, Flatiron Institute, 162 5th Avenue, New York, NY 10010, USA}
	
	\date{\today}
	
\begin{abstract}

We numerically solve the Hubbard model on the Bethe lattice
with finite coordination number \mbox{$z=3$}, 
and determine its zero-temperature phase diagram. For this purpose, we introduce and develop the `variational uniform tree state' (VUTS) algorithm, 
a tensor network algorithm which generalizes the variational
uniform matrix product state algorithm to tree tensor networks. 
Our results reveal an antiferromagnetic insulating phase and a paramagnetic metallic phase, separated by 
a first-order doping-driven metal-insulator transition. 
We show that the metallic state is a Fermi liquid with
coherent quasiparticle excitations for all values of the interaction
strength $U$, and we obtain the finite quasiparticle weight $Z$ from the 
single-particle occupation function of a generalized ``momentum" variable.
We find that $Z$ decreases with increasing $U$, ultimately saturating
to a non-zero, doping-dependent value. Our work demonstrates that tensor-network calculations
on tree lattices, and the VUTS algorithm in particular, are a platform for obtaining controlled
results for phenomena absent in one dimension,
such as Fermi liquids, while avoiding computational difficulties
associated with tensor networks in two dimensions. We envision that future 
studies could observe non-Fermi liquids, interaction-driven metal-insulator 
transitions, and doped spin liquids using this platform.

\end{abstract}

\maketitle

\section{Introduction}

The Hubbard model is a cornerstone of condensed matter physics.  As a paradigmatic model of strongly-correlated electrons  \cite{ANDERSON1196}, it is simple to formulate yet rich in behavior. In two dimensions (relevant e.g. to cuprate superconductors) observed behaviors include, but are not limited to, antiferromagnetism, unconventional metallic behavior characterized by a pseudogap and deviations from Fermi liquid theory~\cite{tremblay2006review,Alloul2014,Gunnarsson2015,Wu2017,Schafer}, as well as stripe orders~\cite{PhysRevB.93.035126,Zheng1155} closely competing with superconducting states at low temperatures~\cite{PhysRevX.10.031016}. 
But despite decades of effort, a comprehensive understanding of the phase diagram of the two-dimensional Hubbard model has not yet been fully reached.
Therefore any solutions of the Hubbard model, whether obtained analytically or by accurate and controlled numerical techniques, are of great value.

The most reliable and comprehensive solutions of the Hubbard model obtained so far have been mainly in (quasi-) one dimension \cite{PhysRevLett.20.1445} and in infinite dimensions (infinite lattice coordination number)~\cite{PhysRevB.45.6479,PhysRevLett.62.324,RevModPhys.68.13}.
In one dimension, solutions can be obtained by integrability~\cite{1D_Hubbard_book} and bosonisation methods~~\cite{Giamarchi_book}, 
as well as numerically with matrix product state (MPS) tensor network methods \cite{PhysRevLett.69.2863,SCHOLLWOCK201196}. 
The latter can also reliably treat quasi-one-dimensional ladder or cylindrical geometries with a small transverse size. 
In the limit of infinite dimensions the Hubbard model is again numerically tractable due to the fact that dynamical mean-field theory (DMFT) becomes exact \cite{RevModPhys.68.13}. Using this technique, the phase diagram of the infinite-dimensional Hubbard model can be
mapped out and a detailed understanding of the interaction-driven Mott insulator to metal transition 
has been established (for reviews, see Refs.~\cite{RevModPhys.68.13,RevModPhys.78.865,
DMFT@25,georges_lectures,rozenberg_lectures}).

While they provide useful insights into the physics of the two-dimensional Hubbard model, these limiting cases also have peculiarities which limit the generality of the 
conclusions that can be drawn from their study. In the limit of infinite dimensions, the metallic state is found to be a Fermi liquid, with interactions affecting one-particle properties in a local 
(momentum-independent) manner only. Hence, in this limit, the feedback of long-wavelength collective modes or even short-range spatial correlations 
on quasiparticle properties is entirely absent. 
These effects are important, in particular close to a critical point. In two dimensions, they are, for example, responsible 
for the formation of a pseudogap \cite{Preuss1997,Macridin2006,Kyung2006,Gunnarsson2015,Scheurer2018}.
In contrast, in one dimension the low-energy excitations consist {\it only} of bosonic collective modes, associated with charge 
and spin degrees of freedom.
There is no notion of a Fermi liquid and the metallic behavior of the Hubbard model is a Luttinger liquid which lacks coherent quasiparticles 
and displays spin-charge separation~\cite{Giamarchi_book}.  

\begin{figure}
	\includegraphics[scale=0.3]{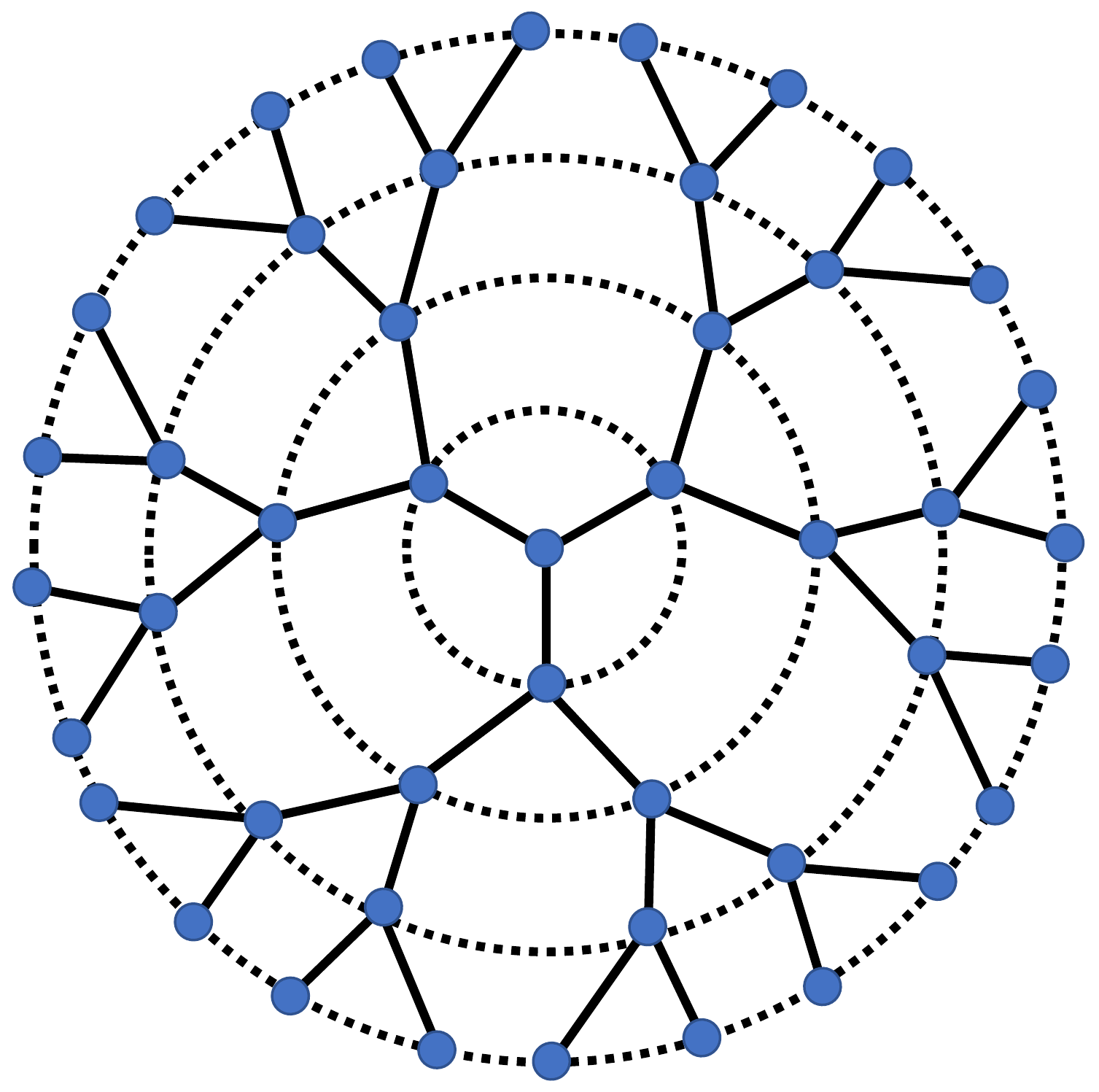}
	\caption{A portion of the infinite Bethe lattice with coordination number $z = 3$. The figure depicts four ``generations" of the tree structure, with a particular site chosen to be the ``center" site. Note that in the actual Bethe lattice there is no special site, and generations can be counted from any of the sites.}
	\label{fig:BL intro}
\end{figure}

In this work we perform a controlled and accurate numerical study of the ground state of the Hubbard model on the Bethe lattice with a finite coordination number $z$, focusing on the case $z=3$. This is an infinite lattice that has a tree structure, where every site is connected to the same number of other sites ($z$) but there are no loops. We show a portion of this lattice in Fig. \ref{fig:BL intro}.
This lattice provides an intermediate case between one dimension ($z=2$) and infinite dimensions ($z = \infty$), with the key virtue that it admits controlled solutions via tensor network methods, including away from half filling and in the presence of strong interactions. 

Exact solutions of models on the Bethe lattice have a long history in statistical mechanics~\cite{Baxter_book}, 
starting with the pioneering article of Hans Bethe~\cite{Bethe_1935}. 
Solutions on the finite coordination number Bethe lattice provide a better approximation to thermodynamic quantities than the mean-field approximation 
(corresponding to the infinite dimensional limit)~\cite{Baxter_book,Bethe_1935,PhysRevLett.74.809}. 
Models studied on the Bethe lattice include classical and quantum spin models~\cite{PhysRevB.80.144415,PhysRevB.77.214431,NAGY2012542,PhysRevB.86.195137,PhysRevB.88.035138,PhysRevB.89.054426,LIU20141}, spin glass systems \cite{Parisi_BL_spin_glass,PhysRevLett.56.1082,doi:10.1002/pssb.200541282,PhysRevB.78.134424}, the Bose Hubbard model \cite{PhysRevB.80.014524}, and models of Anderson localization  \cite{Abou_Chacra_1973,Mirlin_1991,PhysRevLett.78.2803,PhysRevB.100.094201,dupont2020dirty}. The fermionic Hubbard model on the finite version of the $z=3$ Bethe lattice (known as a Cayley tree) has also been studied previously using a variant of the density matrix renormalization group (DMRG) algorithm \cite{Lepetit2000}, but only the case of half filling was studied (which is a charge insulator) and only local ground state quantities given (energy, staggered magnetization and its fluctuations, and neighboring spin correlations). Since that time, there have been significant advances in DMRG and related algorithms for infinite one-dimensional systems, which we generalize here to the Bethe lattice and use to obtain our results. Notably, there has been no previous study of metallic states on the finite connectivity Bethe lattice, to the best of our knowledge. 

We determine the full phase diagram of the fermionic Hubbard model on the $z=3$ Bethe lattice, allowing for a two-site unit cell, and establish 
the nature of the doping-driven Mott insulator to metal transition (MIT). We find that this transition is first-order, and that for every value of interaction strength there is a region of forbidden density. Therefore, in the interaction-density plane the model exhibits phase separation at low doping levels.  
We find that, for all allowed values of the density, the doped metallic ground-state does not display magnetic long-range order. 

Importantly, we also demonstrate that the doped metal hosts coherent quasiparticles at all studied values of the interaction strength, $U$, from weak to strong coupling, and determine the behavior of the quasiparticle weight as a function of $U$. This answers in the affirmative the question of whether Fermi liquid behaviour 
applies as soon as the peculiar kinematic constraints of one dimension are alleviated, and also provides a concrete description of a Fermi liquid ground state with tensor networks, both of which are key motivations for our work. Generally, it is difficult for tensor networks to accurately describe interacting metallic states above one dimension, although much progress has been made in this direction \cite{PhysRevB.81.165104,PhysRevB.93.045116,PhysRevB.100.195141,mortier2020resolving}. Our work provides an alternate route to this agenda, avoiding the computational challenges of two dimensional tensor networks, while going beyond the restrictions of one dimensional physics. 

We obtain our results by generalizing a recently developed MPS method, variational uniform MPS (VUMPS) \cite{PhysRevB.97.045145}, to tree tensor network (TTN) states \cite{PhysRevA.74.022320,PhysRevB.82.205105,doi:10.1063/1.4798639}, which we dub the variational uniform tree state algorithm (VUTS). We further introduce the fermionic version of the VUTS algorithm, using the swap-gate method of Refs. \cite{PhysRevB.81.165104,Orus_review}. The VUTS algorithm works directly in the thermodynamic limit, which is important in the study of models on the Bethe lattice. The alternative is to study the finite Cayley tree and perform a finite-size scaling analysis. However, the number of boundary sites on the Cayley tree is always more than half of the total, and therefore finite-size effects are unusually strong and can even lead to conclusions that do not hold on the Bethe lattice \cite{Baxter_book,PhysRevB.87.085107,OSTILLI20123417}. Working directly in the infinite-size limit is therefore important for models on trees. All previous works studying quantum models on the (infinite) Bethe lattice using tensor networks have used a variant of the infinite time-evolving block decimation (iTEBD) algorithm \cite{PhysRevLett.98.070201}. In the one-dimensional case, the VUMPS algorithm has been found to be much more efficient than other methods that work in the thermodynamic limit, such as iTEBD or earlier infinite DMRG algorithms \cite{PhysRevB.97.045145}, and indeed we find its extension in the form of the VUTS algorithm we develop to be very efficient. Our method scales as $\O(\chi^{z+1})$ where $\chi$ is the bond dimension of the tensor network being optimized. For $z=3$ this scaling is significantly better than that of the most modern and accurate projected entangled pair states (PEPS) algorithms that scale as $\O(\chi^{10})$ for PEPS bond dimension $\chi$ (assuming the boundary MPS bond dimension scales as $\chi^2$) \cite{PhysRevB.92.035142,PhysRevB.94.035133,PhysRevB.94.155123,PhysRevB.98.235148,PhysRevX.9.031041,PhysRevB.100.195141}, but is more challenging than the $\O(\chi^3)$ scaling of DMRG. However, the steeper scaling is mitigated by the fact that the typical bond dimension required to reach an accurate solution generally decreases as one goes to higher dimensions and larger coordination numbers due to the monogamy of entanglement and more mean-field-like properties of the wavefunction.
The accuracy of tensor network methods is often measured by the \emph{truncation error}, which measures the typical loss of fidelity incurred during the truncation step of the optimization algorithm. In this work, for a bond dimension of $\chi = 100$, we are able to achieve a truncation error of less than $10^{-3}$ in the most computationally challenging part of the phase diagram (most entangled ground state), and a truncation error of less than $10^{-7}$ in the best cases. This level of accuracy allows us to measure long-distance correlation functions well enough to extract information about quasiparticle coherence in the metallic phase, which demonstrates that our method can be used to reliably study critical phases of matter on the Bethe lattice.

This work suggests promising
future directions for studying the behavior of strongly correlated electrons in a controlled setup. Because Fermi liquid behavior is a rather generic feature of metallic states, the present study allows to establish a controlled platform which can be used to study how Fermi liquid behaviour can be broken by further perturbations to the model considered here, or in other fermionic models. In the concluding section of this article, we discuss possible routes towards achieving this goal. If successful, tensor network solutions of correlated electrons on the $z=3$ Bethe lattice could provide a new platform for studying non-Fermi liquids~\cite{RevModPhys.73.797,doi:10.1146/annurev-conmatphys-031016-025531} in a controlled and accurate manner. Other potential applications are the study of the interaction-driven MIT in frustrated fermionic systems, and the study of fermions on closely related tree-like lattices, such as the Husimi cactus on which the Heisenberg model and other spin models have been shown to display spin liquid phases~\cite{Chandra_1994}. We elaborate on all these directions and others at the end of the paper.

This paper is organized as follows. In section \ref{sec:VUTS} we describe the general VUTS method, applicable to generic Hamiltonians. In section \ref{sec: Hubbard model} we define the Hubbard model on the Bethe lattice and show the phase diagram obtained from the VUTS solution. Section~\ref{sec:quasiparticles} discusses the calculation of the quasiparticle weight from the occupation function, as well as Luttinger's theorem. Finally, in Sec. \ref{sec:discussion} we summarize and discuss future directions.

\section{Variational uniform tree state algorithm}
\label{sec:VUTS}

\begin{figure}
	\includegraphics[scale=0.25]{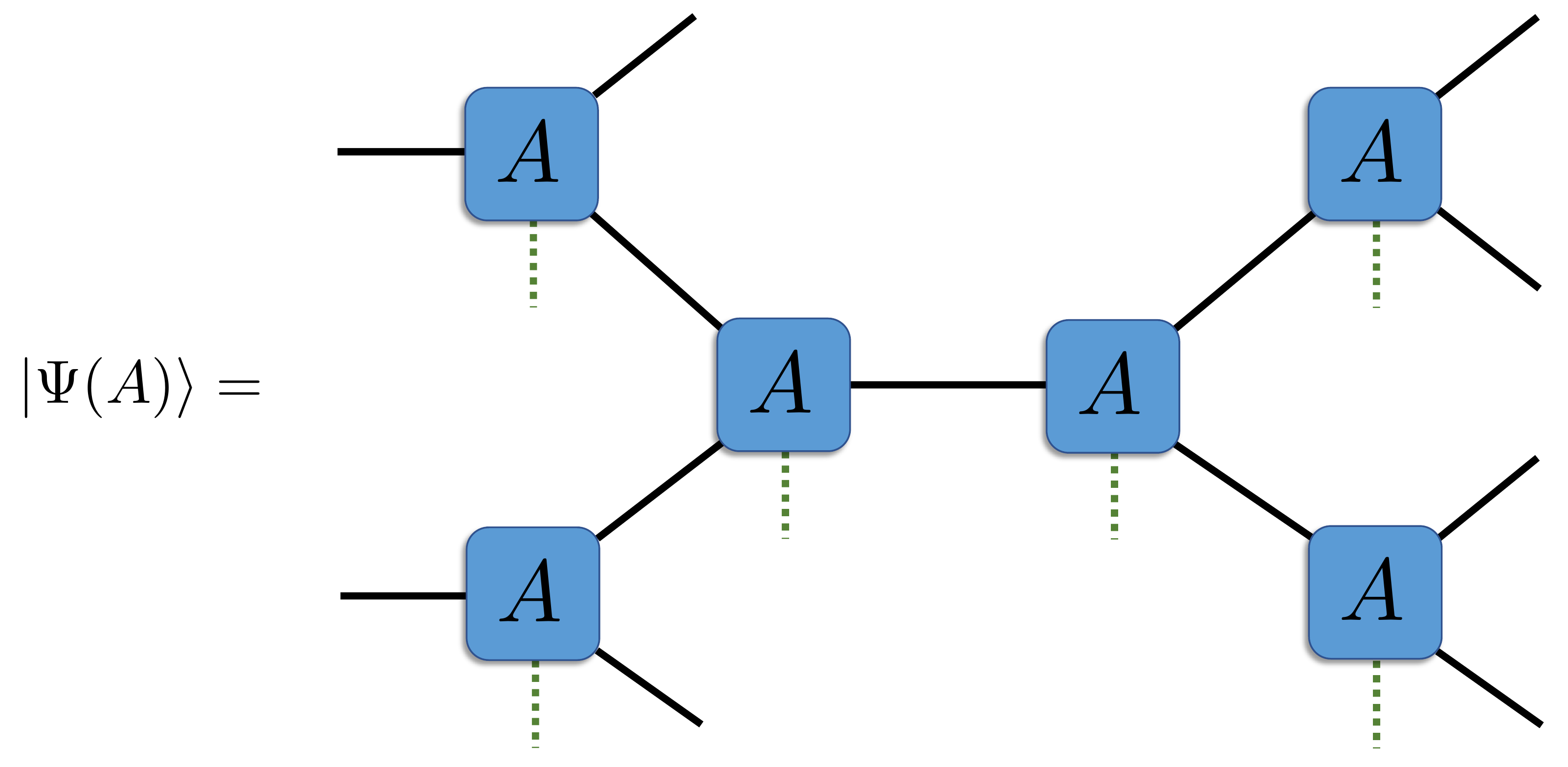}
	\caption{A finite portion of the infinite tree tensor network (TTN) state describing the many-body wave function on the Bethe lattice. The physical legs (green dashed lines) form the nodes of the lattice, while the virtual legs (straight black lines) form the edges. In this case, a single tensor $A$ comprises the state, and the unit cell is just a single site.}
	\label{fig:TTN Bethe lattice unlabeled}
\end{figure}

\begin{figure}
    \includegraphics[scale=0.25]{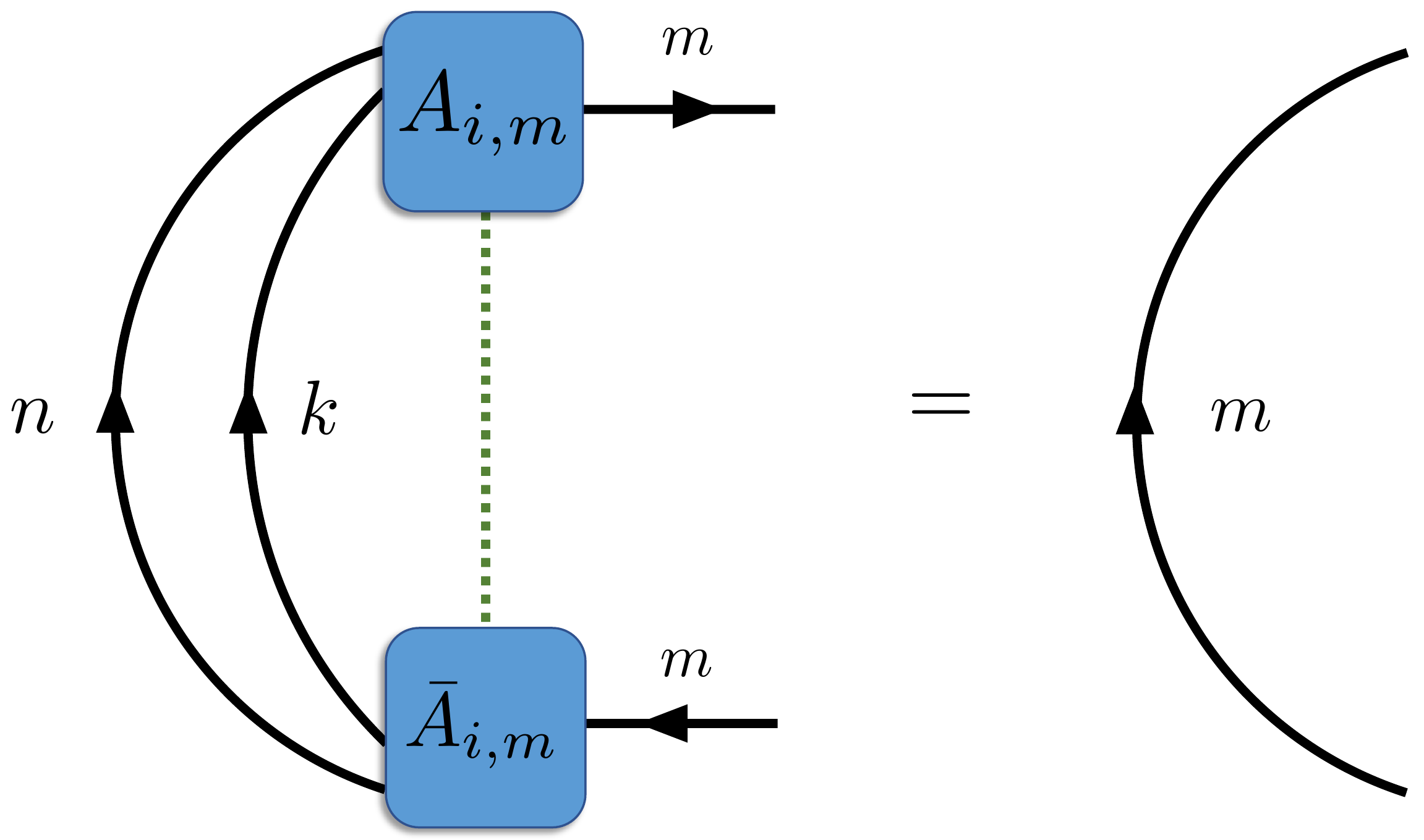}
    \caption{Diagrammatic version of the gauge conditions of Eq. (\ref{eq:canonical gauge}). The bonds labelled $k,n,m$ have link indices $l_k,l_n,l_m$ respectively (and the uncontracted $m$ bond on the ket has link index $l_m'$). The unlabelled dashed bond is the physical degree of freedom with index $s_i$.}
	\label{fig:canonical gauge}
\end{figure}

\begin{figure}
	\includegraphics[scale=0.35]{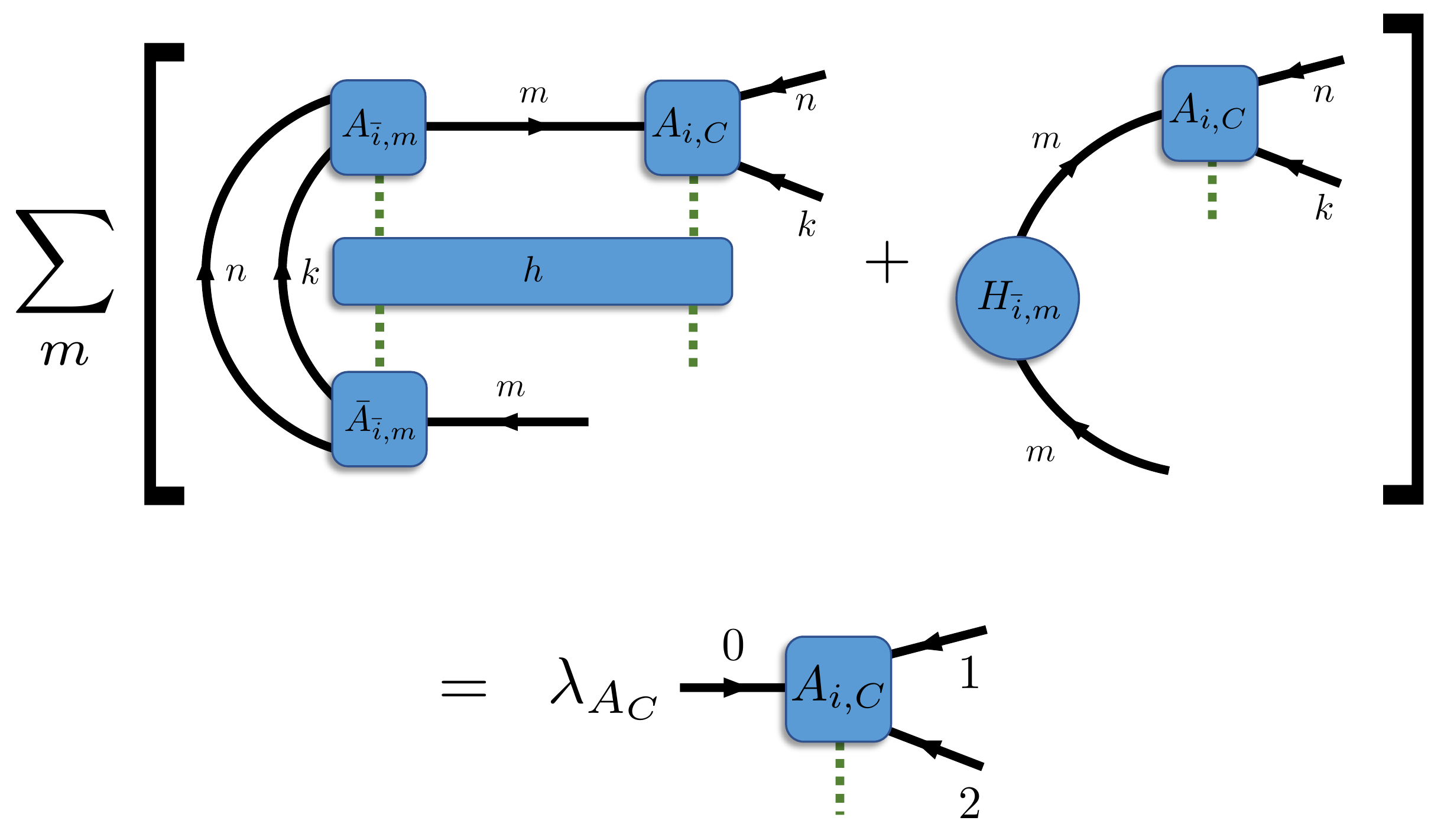}
	\caption{The eigenvalue equation for $A_{i,C}$. The sum over $m$ is a sum over the contributions from each leg of the $A_{i,C}$ tensor.}
	\label{fig:H_Ac}
\end{figure}

\begin{figure}
	\includegraphics[scale=0.45]{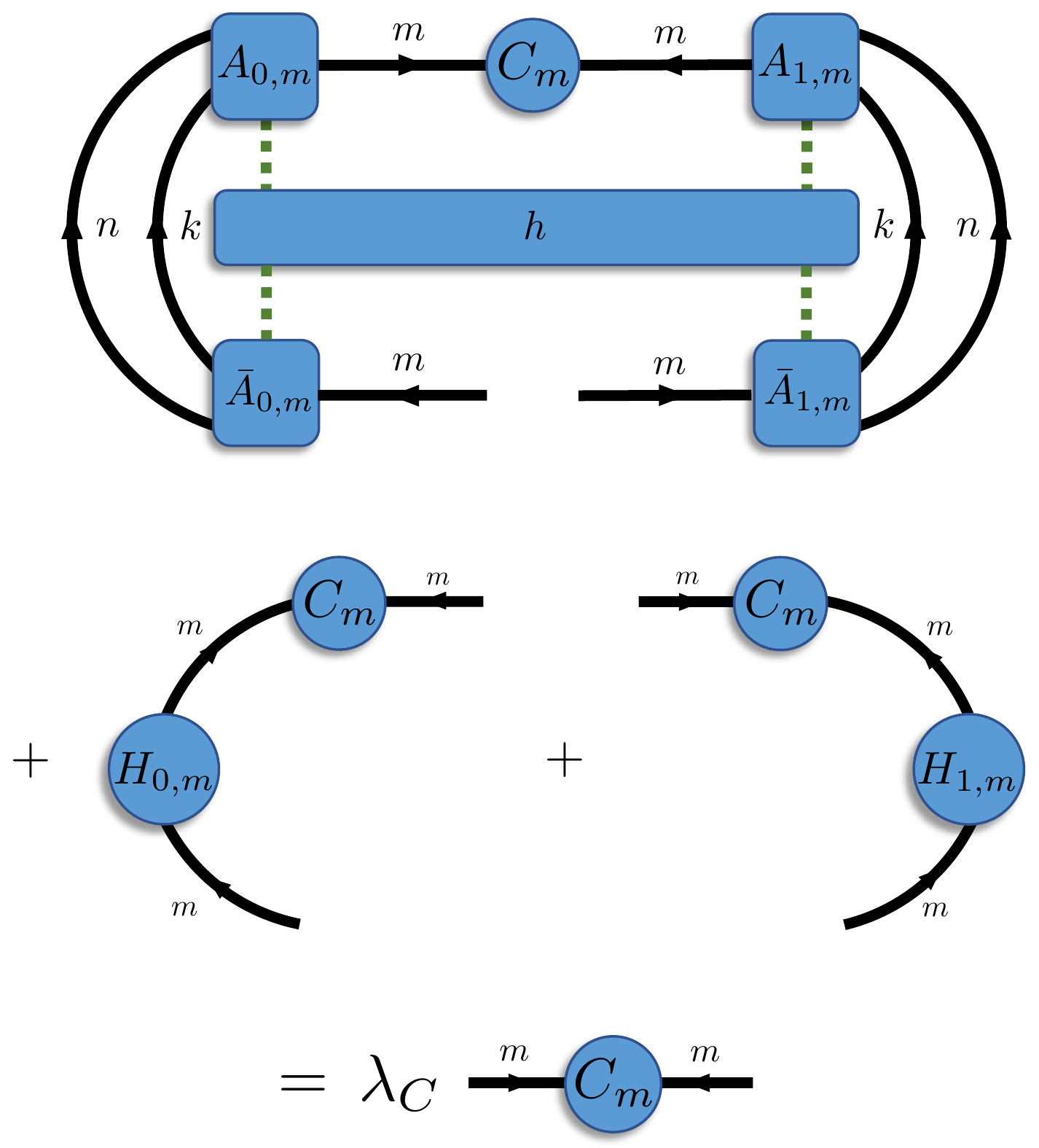}
	\caption{The eigenvalue equation for $C_{m}$.}
	\label{fig:H_C}
\end{figure}

In this section we introduce the variational uniform tree state algorithm (VUTS), a generalization of the variational uniform matrix product state algorithm (VUMPS) \cite{PhysRevB.97.045145}, for optimizing infinite tree tensor network (TTN) states. We start with a Bethe lattice of quantum degrees of freedom. For simplicity we focus on the algorithm for coordination number $z = 3$, which is the value for the model studied in this paper, but the extension to general $z$ is straightforward. We use an infinite TTN as our ansatz to approximate quantum states on the Bethe lattice. For $z = 3$, the infinite TTN with a 1-site unit cell consists of an order 4 tensor $A  \in \mathbb{C}^{\chi \times \chi \times \chi \times d}$, with one physical leg ($s$) which runs over the physical degrees of freedom $1,...,d$, and three virtual legs ($l_0,l_1,l_2)$ that run over virtual degrees of freedom $1,...,\chi$. The virtual legs of neighboring tensors connect to each other, forming the same geometry as the Bethe lattice, as shown in the tensor network diagram in Fig. \ref{fig:TTN Bethe lattice unlabeled}.
\\
\indent 
The Hamiltonians we will focus on here are isotropic and have an equivalence between all sites (the analog of translational invariance for hypercubic lattices). The ground state will potentially break this isotropy completely, and break the site equivalence down to a non-trivial unit cell. In this paper, we allow for the state to be fully anisotropic between the different directions emanating from a given site, but for simplicity we focus exclusively on the case when the unit cell consists of no more than two sites (generalizing to arbitrary unit cells is straightforward). The infinite TTN state we study therefore has a 2-site unit cell, and is parameterized by a set of $2z$ tensors $A_{i,m}$, where $i = 0,1$ labels the location in the unit cell, and $m = 1, \dots, z$ labels the direction of the gauge (defined below). Again, each tensor has one physical index $s_i = 1, \dots, d$ and three virtual ``link" indices $l_0,l_1,l_2$.
\\
\indent
Because the TTN has no loops, it is straightforward to work in the \emph{canonical gauge}, i.e. the gauge where the tensors are constrained to be orthonormal bases when viewed as a matrix from two link indices $l_n,l_k$ and the physical index $s_i$ to the remaining link index $l_m$. This constraint on the tensors is very useful for making the variational optimization faster and more stable, and is standard in a wide variety of tensor network algorithms, particularly in 1D algorithms like VUMPS and DMRG. The constraint on the tensors is written as
\begin{equation}
\displaystyle\sum_{s_i,l_n,l_k} \bar{A}^{s_i,l'_m,l_n,l_k}_{i,m} A^{s_i,l_m,l_n,l_k}_{i,m} = \mathbb{1}^{l'_m,l_m}_{i,m},
\label{eq:canonical gauge}
\end{equation}
where we have introduced the notation $\bar 0 = 1, \bar 1 = 0$ for the unit cell indices and use $\bar{A}$ to denote the complex conjugation of $A$. The matrices $\mathbb{1}_{i,m}$ are identities.  Diagrammatically, Eq. (\ref{eq:canonical gauge}) is equivalent to Fig.~\ref{fig:canonical gauge}. The arrows on the links denote the gauge of the $A$ tensors (the outgoing link is the direction of the gauge). Any TTN can be brought into the form where the tensors obey Eq. (\ref{eq:canonical gauge}) (or equivalently Fig. \ref{fig:canonical gauge}) by inserting a particular set of ``gauge transformations"
onto the link degrees of freedom, i.e. inserting a particular set of resolutions of the identity $X X^{-1}$ (where $X$ is an invertible matrix) onto the links of the TTN. Note that the gauge transformation does not affect the observables of the system, and any TTN can be transformed into the canonical gauge efficiently.
\begin{figure}
	\includegraphics[scale=0.45]{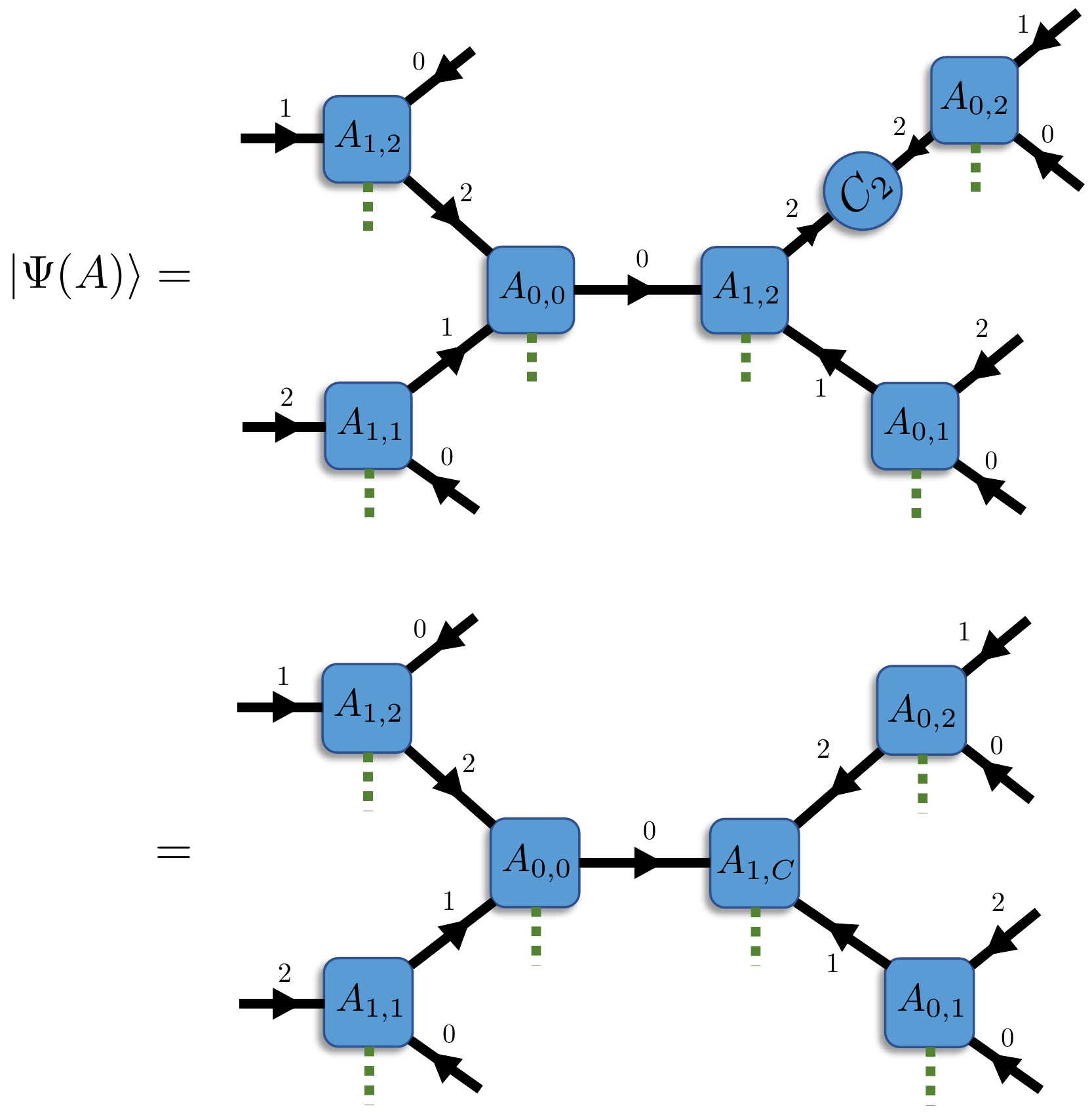}
	\caption{The same portion of the Bethe lattice as in Fig. \ref{fig:TTN Bethe lattice unlabeled}, but with the bonds labeled with $m = 0,1,2$, which have link indices $l_0,l_1,l_2$ respectively. Additionally, each tensor is labeled with subscripts $i,m$, where $i = 0,1$ is the unit cell index and $m$ is the direction of the gauge. In the top diagram, the gauge center $C_2$ is shown on a bond labelled by $2$. The next equality shows that the $C_2$ tensor can be absorbed into the $A_{1,2}$ tensor to put the gauge center on the site tensor, creating $A_{1,C}$.}
	\label{fig:TTN Bethe lattice with gauge centers}
\end{figure}
\\
\indent
Examples of the TTN state with a 2-site unit cell in the canonical gauge are shown in Fig. \ref{fig:TTN Bethe lattice with gauge centers}. In the top diagram, the gauge center is $C_2$. Here, $C_2$ represents the projection of the infinite wavefunction of the system onto the finite-sized Hilbert space of that link of the network. The center matrices $C_m$ constitute invertible gauge transformations relating the $A$ tensors to each other via $A_{i,m} C_m = A_{i,n} C_{n}$, where $i = 0,1$ and $n \neq m$. The center matrices $C_m$ also contain important information like the entanglement spectrum between the two infinite halves of the system split by that link. Additionally, the gauge center can be absorbed onto an $A$ tensor, defining the center site tensors $A_{i,C} = A_{i,m} C_m$ for any $m$. This is shown in the lower diagram of Fig. \ref{fig:TTN Bethe lattice with gauge centers}. In this case, $A_{i,C}$ would represent the infinite wavefunction of the system projected onto a single site (and again, different spectra of that tensor relate to entanglement spectra of different bipartitions of the lattice).
\\
\indent
We describe the algorithm for the case of $H = \sum_{\<i,j\>} h_{i,j}$, where $h_{i,j}$ is a two-site operator that acts on nearest neighbors only. The case of longer-range operators is  treatable using techniques like those described in Appendix~C of Ref. \cite{PhysRevB.97.045145}. The total energy is given by $E = \sum_{\<i,j\>} \<\psi|h_{i,j}|\psi\>$, and we want to minimize $E$, treating the tensor elements of our TTN as variational parameters. As in VUMPS (and many related tensor network ground state methods), VUTS proceeds in three main steps that are iterated until convergence:
\begin{enumerate}
  \item Compute the projected Hamiltonians (the Hamiltonian projected into the basis corresponding to the virtual degrees of freedom of the network) to turn the global optimization into a local optimization problem.
  \label{alg:projected Hamiltonian}
  \item Find the optimized tensors by minimizing the energy of the projected Hamiltonian.
  \label{alg:optimize tensors}
  \item Update the tensor network with the new optimized tensors.
  \label{alg:update network}
\end{enumerate}
\indent
To begin, say we are interested in optimizing a 1-site projected wavefunction $A_{i,C}$, as defined in Fig. \ref{fig:TTN Bethe lattice with gauge centers}. Step \ref{alg:projected Hamiltonian}  requires computing infinite sums of local Hamiltonian terms, projected into the basis of our gauged TTN (defined by the tensors $A_{i,m}$), for each of the $z=3$ infinite subtrees connected to $A_{i,C}$. In order to perform the infinite sum, we focus on summing the energy contributions of a single subtree. An example for the series that needs to be summed for the $m=2$ direction in order to optimize the $A_{0,C}$ tensor is shown in Fig. \ref{fig:H_{1,2} series}. We define the results of these summations as the matrices $H_{i,m}$. The summation can be carried out by making use of the fact that the sum is a geometric series. However, care has to be taken to project out infinite energy contributions to keep the series convergent (i.e. keep the norm of the solution $H_{i,m}$ from diverging). The procedure of performing the summation and projecting out the infinite energy contributions is a generalization of the one in Appendix D of Ref. \cite{PhysRevB.97.045145}, and we discuss it in more detail in Appendix \ref{subsec:summing Hamiltonian terms}.
\begin{figure}
	\begin{subfigure}[b]{0.45\textwidth}
          \includegraphics[width=1\linewidth]{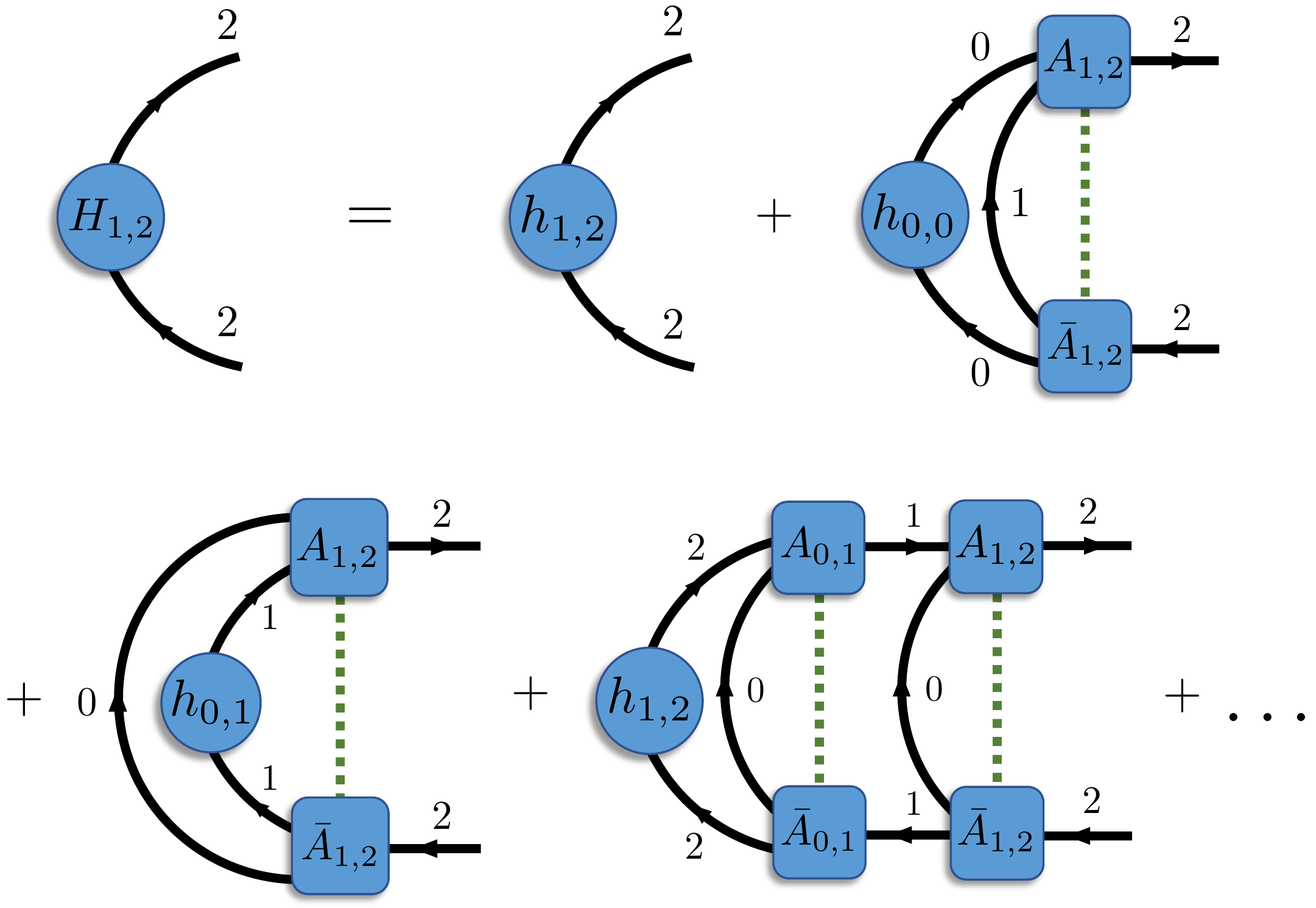}
          \caption{}
        \end{subfigure}
        \begin{subfigure}[b]{0.38\textwidth}
          \includegraphics[width=1\linewidth]{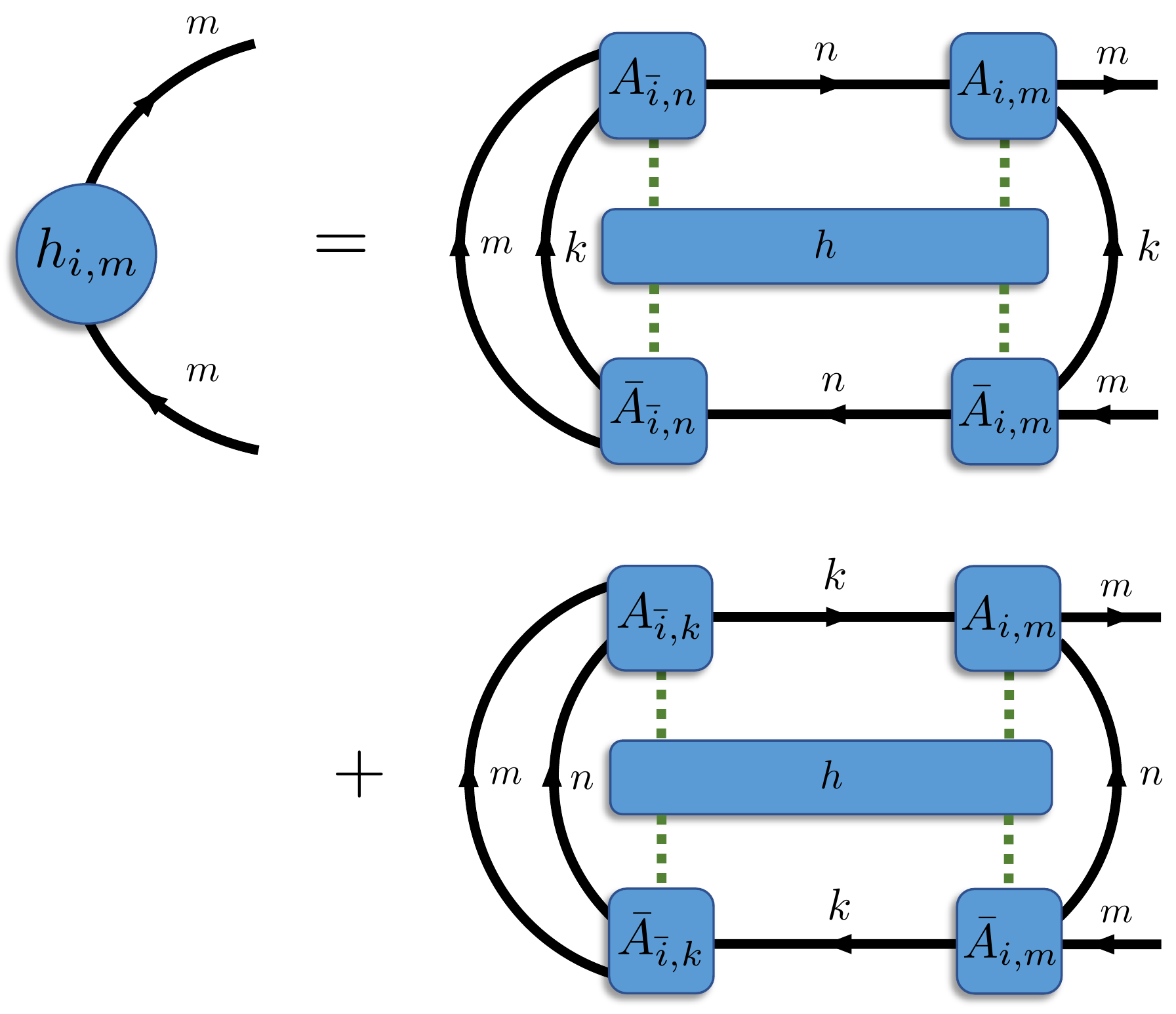}
          \caption{}
          \label{fig:h_i_m}
        \end{subfigure}
	\caption{(a) The first few terms of the series for $H_{1,2}$, the projected Hamiltonian contribution for one of the three branches of the infinite Bethe lattice connected to the center site tensor $A_{0,C}$ (on the $i=0$ sublattice). (b) Definition of the $h_{i,m}$ tensors used in (a), which are a sum of two local Hamiltonian environment tensors sitting on two branches of the Bethe lattice. Note that the tensor labeled $h$ represents the two-site operator term $h_{i,j}$, which the Hamiltonian is made up of.}
	\label{fig:H_{1,2} series}
\end{figure}
\\
\indent 
Once the environment tensors are found, we can proceed to step \ref{alg:optimize tensors} of the algorithm, which we begin by optimizing $A_{i,C}$. This is done by finding the ground state of the Hamiltonian projected onto the sublattice site $i$, a standard procedure in VUMPS and DMRG. The eigenvalue equation for $A_{i,C}$ is shown diagrammatically in Fig. \ref{fig:H_Ac}, and is solved iteratively (using a Hermitian eigensolver such as Lanczos). To find the TTN ground state, we obtain the eigenvector with the smallest eigenvalue. As in the VUMPS algorithm, in addition to optimizing $A_{i,C}$, we also optimize $C_m$. $A_{i,C}$ and $C_m$ are then used to solve for $A_{i,m}$, which make up the updated infinite TTN state (see next paragraph). The eigenvalue equation for $C_m$ is shown diagrammatically in Fig. \ref{fig:H_C}.
\\
\indent
Finally, once $A_{i,C}, C_m$ for all $i = 0,1, m = 0,1,2$ are optimized, we can proceed to step \ref{alg:update network} of the algorithm and solve for our new $A_{i,m}$ tensors by minimizing
\begin{equation}
\epsilon_{i,m}
=
\min_{A_{i,m}^{\dagger} A_{i,m} \, = \, \mathbb{1}_{i,m}} || A_{i,C} - A_{i,m} C_m ||. 
\label{eq:new A tensors}
\end{equation}
This minimization problem can be solved optimally using techniques described in Eqs. (18)-(22) of Ref. \cite{PhysRevB.97.045145}. The new $A_{i,m}$ we obtain constitute our updated TTN, and steps \ref{alg:projected Hamiltonian}-\ref{alg:update network} are repeated until convergence. Convergence is achieved when the largest error found in Eq. (\ref{eq:new A tensors}), $\epsilon_{\text{prec}} \equiv \max\{\epsilon_{i,m}\}$, falls below a chosen threshold (e.g. $\epsilon_{\text{prec}} < 10^{-12}$).
\\
\indent
The VUTS algorithm with a 1-site update, as we describe here, scales as $O(\chi^{z+1})$  which becomes $O(\chi^4)$ for the $z=3$ Bethe lattice and $O(\chi^3)$ for $z=2$, thus reducing to the scaling of VUMPS in the $z=2$ case. Additionally, a 2-site update can be formulated, analogous to the 2-site DMRG algorithm which is commonly used. This requires a slight modification of the algorithm where ground states of 2-site and 1-site projected Hamiltonians are computed (as opposed to 1-site and 0-site projected Hamiltonians in the version of the algorithm described above). This can lead to improved convergence since a larger local Hilbert space is explored, but has a higher computational cost of $\O(\chi^5)$ for $z=3$. We use this technique at lower bond dimensions in more challenging parts of the phase diagram (near the phase transition), and switch to the 1-site algorithm later in the calculation to reach higher bond dimensions. To dynamically change the bond dimensions, we use a generalization of the subspace expansion procedure described in Appendix B of Ref. \cite{PhysRevB.97.045145}.
\\
\indent 
For fermionic models like the Hubbard model, we need to use a fermionic version of VUTS. We use the method outlined in Refs. \cite{PhysRevB.81.165104,Orus_review}. Every tensor is now endowed with a fermion parity $Z_2$ quantum number and is parity-preserving. When two tensor legs cross on a planar projection of a tensor diagram, a fermionic swap gate is placed at the crossing. In order to employ this method, we need to use a fixed ordering convention for the legs of the tensors $A_{i,m}, A_{i,C}, C_m$, which must be kept consistent in all of the diagrams in the calculation. We address the associated subtleties and details of this approach in Appendix \ref{subsec:fermions}. Other symmetries beyond the $Z_2$ parity can also be used, such as $U(1)$ particle number conservation (to fix the filling), $U(1)$ spin projection symmetry in the z-direction, and also the spin $SU(2)$ symmetry. The inclusion of these symmetries makes the tensors more sparse and should therefore make the tensor operations more efficient, allowing us to reach larger bond dimensions. In this work, we only employ parity quantum numbers, and leave the use of additional symmetries for future work.

\section{Model and phase diagram} 
\label{sec: Hubbard model}

\begin{figure}
	\includegraphics[scale=0.8]{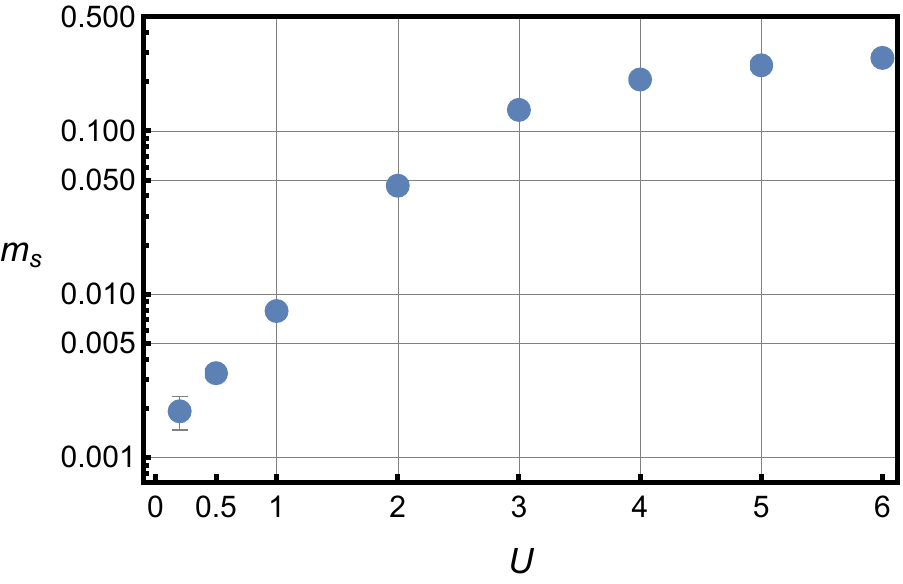}
	\caption{The extrapolated values of the staggered magnetization $m_s$ of the insulating state at half filling, plotted as a function of $U$. Note the logarithmic scale used for $m_s$. The error bars are the discrepancy in the extrapolation with and without the last data point (for the points where they are absent they are smaller than the data points).}
	\label{fig:ms extrapolation}
\end{figure}

\begin{figure}
	\begin{subfigure}[c]{0.45\textwidth}
          \includegraphics[width=0.85\linewidth]{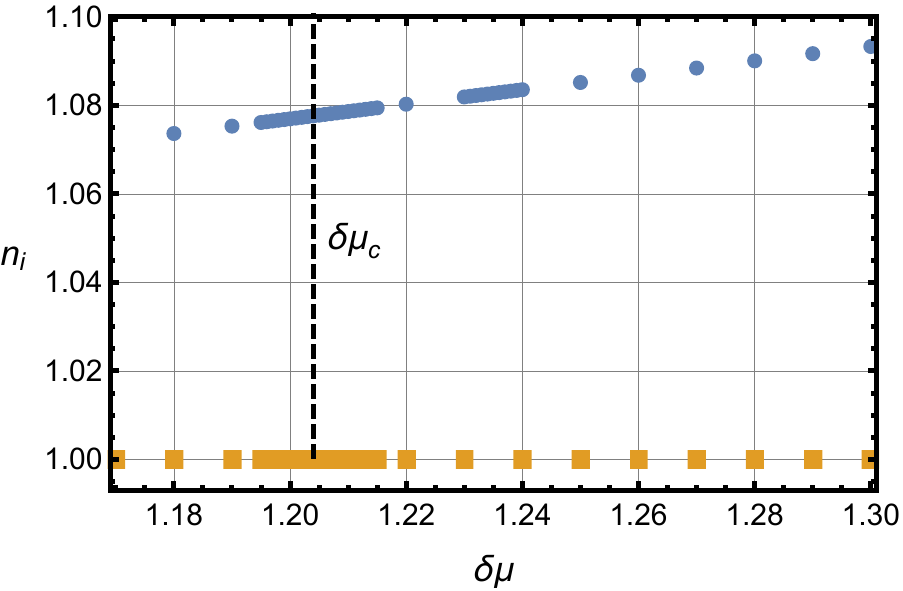}
          \caption{$U = 6, \chi = 50$.}
        \end{subfigure}
	\begin{subfigure}[c]{0.45\textwidth}
          \includegraphics[width=0.85\linewidth]{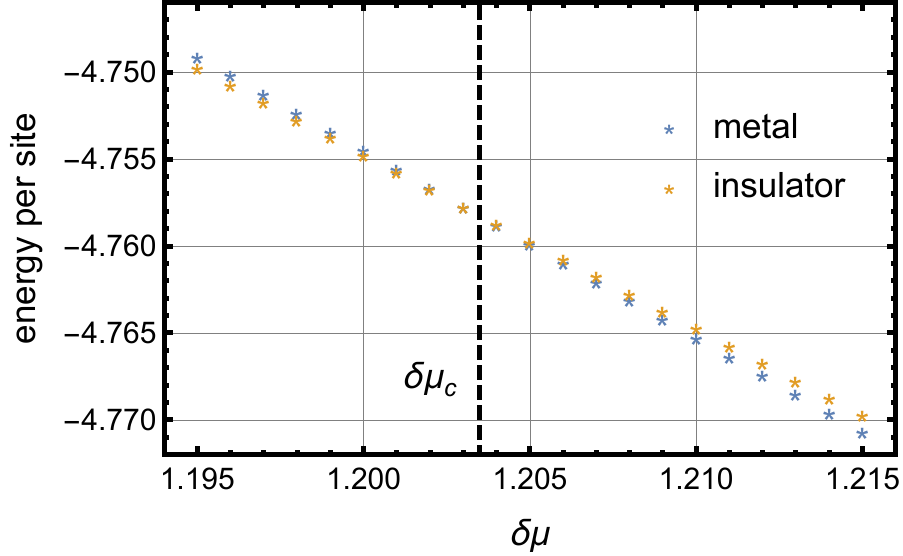}
          \caption{$U = 6, \chi = 50$.}
        \end{subfigure}
	\caption{(a) The occupation versus $\delta \mu$ for $U = 6$ and $\chi = 50$ for the both metallic and insulating branches. All other values of $U,\chi$ look similar. (b) The energy per site of both branches in (a). 
	Their crossing point, $\delta \mu_c$ is indicated by a dashed line in both (a) and (b).}
	\label{fig:n and energy vs delmu for the two branches at U = 6 and chi = 50}
\end{figure}

\begin{figure}
	\includegraphics[scale=0.8]{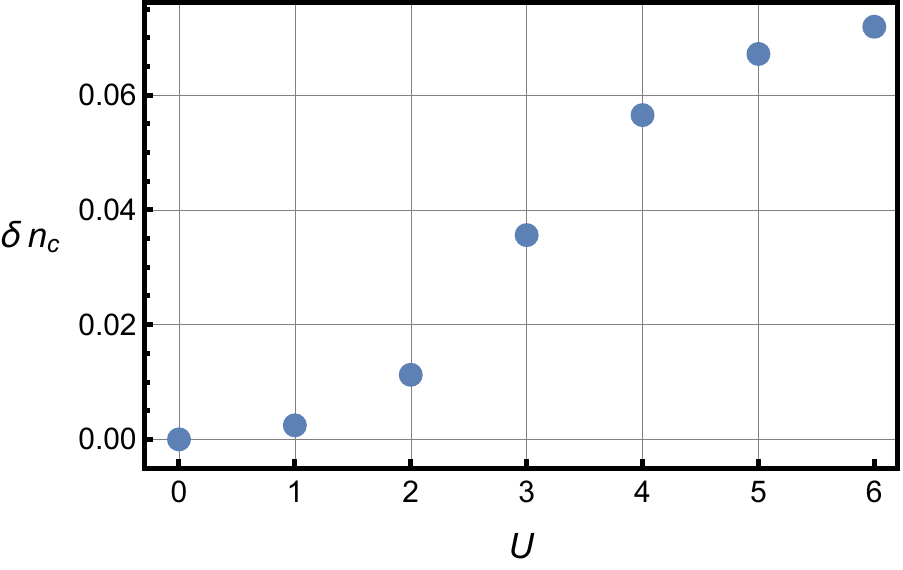}
	\caption{The extrapolated values of the jump in density $\delta n_c$ at the first-order transition occuring at $\delta \mu_c$ plotted as a function of $U$. The error bars, which are the discrepancy in the extrapolation with and without the last data point, are smaller than the data points.}
	\label{fig:jump in occupation extrapolation}
\end{figure}

\begin{figure}
	\includegraphics[scale=0.8]{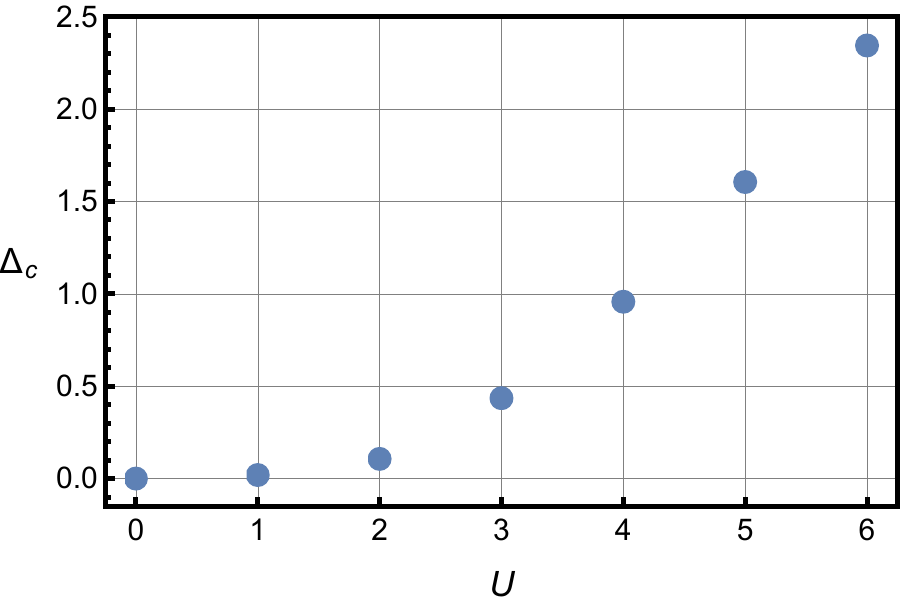}
	\caption{The extrapolated values of the charge gap $\Delta_c$ at half filling plotted as a function of $U$. The error bars, which are the discrepancy in the extrapolation with and without the last data point, are smaller than the data points.}
	\label{fig:charge gap extrapolation}
\end{figure}

\begin{figure}
	\includegraphics[scale=0.8]{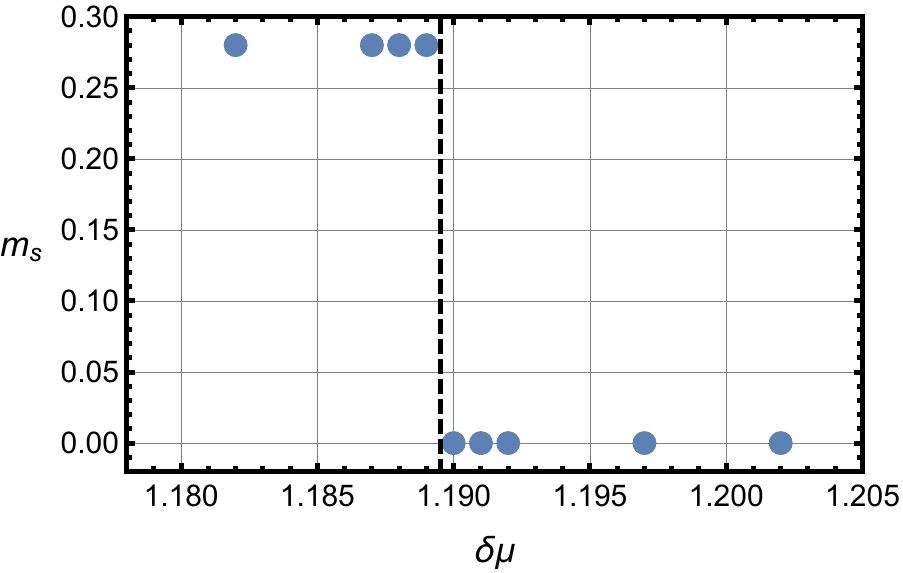}
	\caption{The staggered magnetization $m_s$ versus the chemical potential $\delta \mu$ for $U = 6$ and $\chi = 90$. The critical point $\delta \mu_c$ is indicated by a dashed line. We see the drop from $m_s > 0$ to $m_s = 0$, indicating the first-order transition from the antiferromagnetic insulator to the paramagnetic metal.}
	\label{fig:ms of metallic phase}
\end{figure}

\begin{figure}
	\begin{subfigure}[]{0.49\textwidth}
          \includegraphics[scale=0.8]{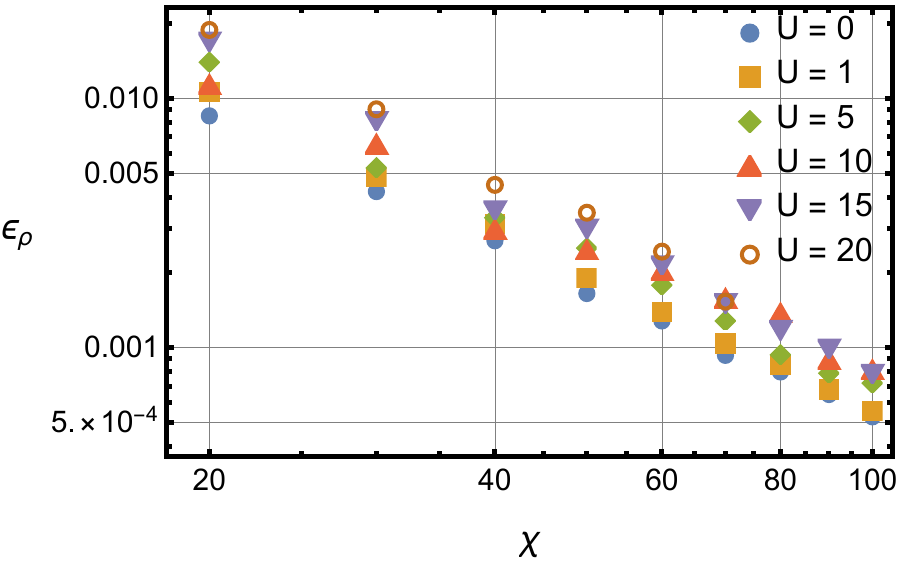}
          \caption{}
        \end{subfigure}\hspace{0.01\textwidth}
	\begin{subfigure}[]{0.49\textwidth}
          \includegraphics[scale=0.8]{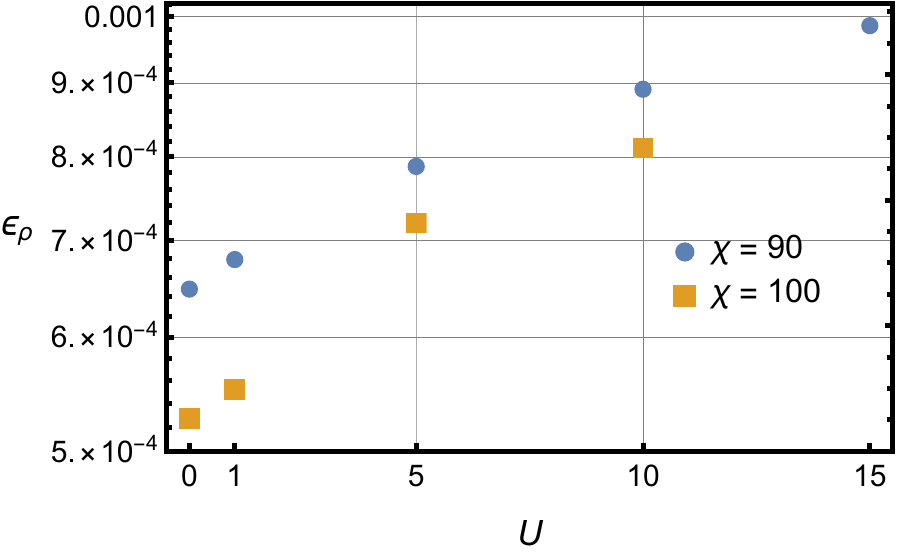}
          \caption{}
        \end{subfigure}
	\caption{Our estimate for the truncation error $\epsilon_\rho$ in the metallic phase for $\<n_i\> = 1.2$ (a) as a function of $\chi$ and (b) as a function of $U$ for the two largest $\chi$. Note that since (a) is a log-log plot the linear form indicates an algebraic relationship.}
	\label{fig:truncation error metal n = 1.2}
\end{figure}

The Hubbard Hamiltonian is given by
\begin{align}
H = - t \sum_{\langle i,j \rangle} \sum_{\sigma = \uparrow, \downarrow}c_{i, \sigma}^{\dagger} c_{j, \sigma}  + U \sum_i n_{i, \uparrow} n_{i, \downarrow}  - \mu \sum_{i, \sigma} n_{i, \sigma},
\label{eq:interacting Hamiltonian}
\end{align}
where the site index $i$ now runs over the Bethe lattice and $n_{i,\s} = c_{i, \sigma}^{\dagger} c_{i, \sigma}$ is the on-site density for an electron of spin $\s$. We set $t = 1$ and vary $U \geq 0$ and $\mu$. Since the model is particle-hole symmetric, we only need to consider $\delta \mu \equiv \mu - \frac{U}{2} \geq 0$. To obtain the phase diagram, we compute the ground state of the Hubbard model using the fermionic VUTS algorithm for several values of the bond dimension, $\chi$. The details of the numerical calculation are given in Appendix \ref{sec:numerical details}. We perform extrapolations in $\chi$, the results of which we describe below. The additional plots of the finite $\chi$ results and details of the extrapolations are in Appendix \ref{sec:phase diagram more plots}. 
\\
\indent 
At half filling, $\delta \mu = 0$, the system is a charge insulator with antiferromagnetic order for all $U > 0$, as in one dimension. To illustrate this we compute the staggered magnetization of the bipartite sublattice, $m_s \equiv \abs{(\< \vec{S}_A \> - \< \vec{S}_B \>)/2}$. The extrapolated values $m_s(\chi\to\infty)$ are shown in Fig. \ref{fig:ms extrapolation}. 
We can see that for any $U>0$ the magnetization is non-zero, tending to zero as $U \rightarrow 0$. From general mean-field theory considerations we expect $m_s \sim e^{-\frac{c}{U}}$ for small $U$. 
However, we did not attempt to confirm this functional form numerically by systematic calculations at very small values of $U$.
\\
\indent 
To illustrate the charge gap at half filling, we compute the on-site occupation $\< n_{i} \> = \sum_{\sigma} \< n_{i,\sigma} \>$ as a function of $\delta \mu$. We show this in Fig. \ref{fig:n and energy vs delmu for the two branches at U = 6 and chi = 50} for a single value of $U = 6$ and $\chi = 50$. All other $U$ and $\chi$ look qualitatively similar.
We see that there are two branches of VUTS solutions: insulating ($\< n_{i} \> = 1$) and metallic ($\< n_{i} \> > 1$). The insulating branch exists for $\delta\mu\leq\delta\mu_1$, and the metallic branch for $\delta\mu\geq\delta\mu_2$: these 
two values, $\delta\mu_{1}$ and $\delta\mu_{2}$ (with $\delta\mu_{2} < \delta\mu_{1}$), are spinodal values limiting the meta-stability of the insulating and metallic solutions, respectively (see Appendix \ref{sec:phase diagram more plots} for details).
For each value of $\delta \mu$ the ground state is the branch with the lower energy. The energies of the two branches cross at a particular value, which we define to be $\delta\mu_c (\chi)$. At $\delta\mu_c (\chi)$, the ground state changes from insulating for $\delta\mu < \delta \mu_c$ to metallic for $\delta\mu > \delta \mu_c$. The occupation undergoes a finite jump, $\delta n(\delta \mu_c, \chi) = \< n_i(\delta \mu_c, \chi) \> - 1$, indicating that this is a first order metal-insulator transition. We estimate the size of the jump in the real system by the $\chi \to \infty$ extrapolated values, which are shown in Fig. \ref{fig:jump in occupation extrapolation}. 
We can see they remain finite for all $U$, meaning that this is a true first-order transition, and not an artifact of finite bond dimension. Using a derivation based on the Maxwell construction (detailed in Appendix~\ref{sec:phase diagram more plots}), 
it can be shown that the total charge gap
is given by $\Delta_c (\chi) = 2 \, \delta\mu_c (\chi)$. In order to obtain the charge gap $\Delta_c$ for the real system, we extrapolate $\Delta_c (\chi)$ in $\chi$. The result is shown in Fig. \ref{fig:charge gap extrapolation}.
We can see that the extrapolated gap shows an exponential-like behavior at small $U$ similar to the one-dimensional case, though the precise behavior is difficult to extract numerically. At large $U$ the gap crosses over to a more linear dependence on $U$. 
\\
\indent 
The first-order transition we observe implies that for every $U$ there is a range of forbidden density. Hence, in the $(U,n)$ plane the model exhibits phase separation. Our numerical method works in the grand canonical ensemble (we do not fix particle number per unit cell), so we cannot observe this phase separation directly. However, VUTS, similar to other variational tensor network methods like DMRG and VUMPS, can get ``stuck" in local minima. We use this fact to find both branches of solutions near the transition point, even when they are meta-stable (i.e. not the lowest energy states), by way of hysteresis in the numerical algorithm (see Appendix \ref{sec:phase diagram more plots} for a detailed explanation). The resulting branches, shown in Fig. \ref{fig:n and energy vs delmu for the two branches at U = 6 and chi = 50}, can be continued to find the spinodal values of the first-order transition (see Appendix \ref{sec:phase diagram more plots}).
\\
\indent 
The metallic ground-state has no magnetic order for any value of density: we find that the staggered magnetization vanishes once $\delta \mu$ crosses $\delta \mu_c$. To illustrate the typical magnetization behavior we observe, in Fig. \ref{fig:ms of metallic phase} we show the staggered magnetization $m_s$ as a function of $\delta \mu$ across $\delta \mu_c$ for $U = 6$ at a large but fixed $\chi = 90$.
Once $\delta \mu$ becomes large enough to drive the system metallic, the staggered magnetization $m_s$ immediately drops to a very small value, which is zero within our error tolerance. All other values of $U$ and $\chi$ behave similarly.  As $\chi$ increases, the small value of magnetization in the metal decreases further, although the behavior is not monotonic, as shown in Appendix \ref{sec:phase diagram more plots}. Also, in Appendix \ref{sec:numerical details} we describe the strategy we use to make sure we do not bias the magnetization of the metallic solution with our ansatzes.
\\
\indent 
It is interesting to compare our results on the $z=3$ Bethe lattice to those established for the doping-driven MIT 
in the $z=\infty$ limit where DMFT becomes exact. 
Only a few studies~\cite{camjayi_rozenberg_2006,wang_millis_2009,fratino_tremblay_prb_2017} consider this transition while also taking into account phases with magnetic long-range order. As in our results, an antiferromagnetic insulator is found at half-filling for a range of chemical potentials, as well as a non-magnetic metallic solution which can be stabilized for values of the chemical potential above a spinodal value $\delta \mu_{2}$. 
Furthermore, a magnetic metallic solution is found to exist in a narrow range of chemical potentials, which connects the magnetic insulator and the non-magnetic metal. 
This may appear to differ from our findings, but it should be emphasized that all these studies consider only non-zero temperatures. As temperature is lowered, it is reported in Refs~\cite{camjayi_rozenberg_2006,wang_millis_2009} that the magnetic metallic solution appears to exist only in an increasingly narrow interval of chemical potentials, and Ref.\cite{camjayi_rozenberg_2006} suggested that at low temperature the MIT is a first-order transition between the magnetic insulator and the non-magnetic metal, with a forbidden range of density corresponding to phase separation. Although, to the best of our knowledge, this has not yet been fully established directly at $T=0$ for $z=\infty$, this conclusion is consistent with our findings on the finite coordination number lattice. In contrast, on fully frustrated $z=\infty$ lattices, which do not allow for long-range magnetic order (e.g. on the fully connected lattice with random hopping), it is established that the 
doping-driven MIT is second order at $T=0$ and becomes first-order only at finite temperature~\cite{RevModPhys.68.13,moeller_1995,kotliar_2002,werner_2007}. 
\\
\indent 
To conclude this section, we mention briefly the numerical accuracy of the data presented here. As noted in the Introduction, the numerical accuracy in tensor networks is generally measured by the truncation error, denoted by $\epsilon_{\rho}$. We show here the scaling of $\epsilon_{\rho}$ in the metallic state, which is the state with the largest entanglement and therefore the most computationally challenging for tensor networks. 
In Fig. \ref{fig:truncation error metal n = 1.2} we plot our estimate of $\epsilon_\rho$ at a fixed density of $\<n_i\> = 1.2$ as a function of $\chi$ for various $U$, and also as a function of $U$ at the largest values $\chi = 90,100$.
We can see that the decay with $\chi$ is algebraic, as expected for a gapless system. As a function of $U$, $\epsilon_\rho$ increases initially and then potentially saturates, although the large $U$ behavior is undetermined. Notably, we can see that $\epsilon_\rho < 10^{-3}$ for all $U$ and $\chi = 100$, which is a high level of accuracy. We discuss the behavior of $\epsilon_\rho$ at other points in the phase diagram in Appendix \ref{sec:numerical details}.

\section{Quasiparticles}
\label{sec:quasiparticles}

In this section we address the existence of quasiparticles in the system. We do this by computing the quasiparticle weight $Z$ from the ``momentum" distribution function of the ground state, which in turn is obtained from real-space correlation functions. 
One peculiarity of the Bethe lattice, is that correlations between any two degrees of freedom sitting on individual nodes of the lattice have a maximal finite correlation length, even at criticality, due to the geometry of the lattice. However, algebraically-decaying correlations reappear after a change of basis to single-particle states that are weighted sums of all nodes of a given generation emanating from a chosen center site. These bases of states unveil the traditional long-range criticality present in gapless states on the Bethe lattice. Below, we introduce a subset of these weighted states called the \emph{symmetric} states, which we focus on in the rest of this section. From these symmetric states we define a quantum number which plays an analogous role on the Bethe lattice to quasi-momentum on the hypercubic lattice, despite the absence of conventional translation invariance.

\subsection{Single-particle basis of symmetric states}

For $U = 0$, the free particle Hamiltonian was diagonalized in Ref. \cite{PhysRevB.63.155110}, using the \emph{symmetric} set of single-particle states. These are given as follows. Choose any site to be labeled as the origin, with site label $0$. Then consider all permutations of the nodes at each generation $l$ from the center. The symmetric states are those which are invariant under all such permutations. Their creation operators are given by
\begin{equation}
\tilde{c}^{\dagger}_{0,\s} \; \equiv \; c^{\dag}_{0,\s}
\label{eq:single particle states l 0}
\end{equation}
and 
\begin{equation}
\tilde{c}^{\dagger}_{l,\s} \; \equiv \; \frac{1}{\sqrt{z(z-1)^{l-1}}} \displaystyle\sum_{\eta_1 = 0}^{z-1} \; \displaystyle\sum_{\eta_2 \neq \eta_1 } \dots \displaystyle\sum_{\eta_l \neq \eta_{l-1}} c^{\dag}_{\eta_1 + \eta_2 + \dots + \eta_l,\s}
\label{eq:single particle states}
\end{equation}
for $l > 0$. The collection of $\eta_i$ denotes a unique path from the origin to the $l$-th generation (this is the usual notation for nodes on the Bethe lattice). The state $\tilde{c}^{\dagger}_{l,\s} |\text{vacuum}\>$ is the symmetric combination of all the singly-occupied spin-$\s$ states of the $l$-th generation of the tree. These states form an orthonormal subset of all the states on the Bethe lattice, but for $U = 0$ they are the only relevant ones. In the symmetric state basis, the free particle Hamiltonian maps onto fermions hopping on an infinite half-chain, with the first hopping amplitude equal to $\sqrt{z}$ and all the rest equal to $\sqrt{z-1}$ (remember $t$ has been set to one). The conjugate variable that replaces momentum is an angle $\theta \in \[0,\pi\]$, and the band energy is given by $\ep(\theta) = 2 \sqrt{z-1} \, \cos\theta$. Note that the energy of a regular one-dimensional band is obtained by replacing $\theta$ with a momentum $k$ and $z$ with $2$. The single-particle wavefunctions $\psi_l(\theta)$ that diagonalize the Hamiltonian are given by
\begin{align}
\begin{split}
\psi_0(\theta) &= \sqrt{\frac{2}{\pi}} \frac{\sqrt{z(z-1)}\sin(\theta)}{\sqrt{z^2 - 4(z-1) \cos^2(\theta)}},
\\ 
\psi_{l \neq 0}(\theta) &= \sqrt{\frac{2}{\pi}} \sin(l \cdot \theta + \gamma(z,\theta)),
\\
\gamma(z,\theta) &= 
\begin{cases}
\arcsin\left(\frac{z \sin (\theta )}{\sqrt{z^2-4 (z-1) \cos ^2(\theta )}}\right),        & \hspace{-2mm} \th \in \[0,\frac{\pi}{2}\) \\
\pi -\arcsin\left(\frac{z \sin (\theta )}{\sqrt{z^2-4 (z-1) \cos ^2(\theta)}}\right),        & \hspace{-2mm} \th \in \[\frac{\pi}{2},\pi\].
\end{cases}
\end{split}
\label{eq:real space wavefunctions n U=0 exact}
\end{align}
\\
\indent 
Once $U \neq 0$, the symmetric states are no longer enough to describe the system. Indeed, if we simply consider a Mott-like state with one particle sitting on each of the sites at generation $l = 1$ away from the center site, that state cannot be written using only the single-particle symmetric states associated with the same center site. 
In general, an arbitrary multi-particle Fock state cannot be constructed as a tensor product of the symmetric single-particle states. Therefore, to construct it one must employ states from other symmetry sectors. The interacting ground states we find numerically in this work therefore contain states from various symmetry sectors. However, excitations above the ground state can occur in any of these sectors, and we do not have to consider all of them. In order to tractably answer the question of existence of quasiparticles, we choose to focus on excitations in the symmetric sector. How quasiparticles in different symmetry sectors are related to each other is an interesting question for future work \cite{Eckstein_thanks}.

\subsection{$\th$-distribution function for $U = 0$}

\begin{figure}
	\includegraphics[scale=0.4]{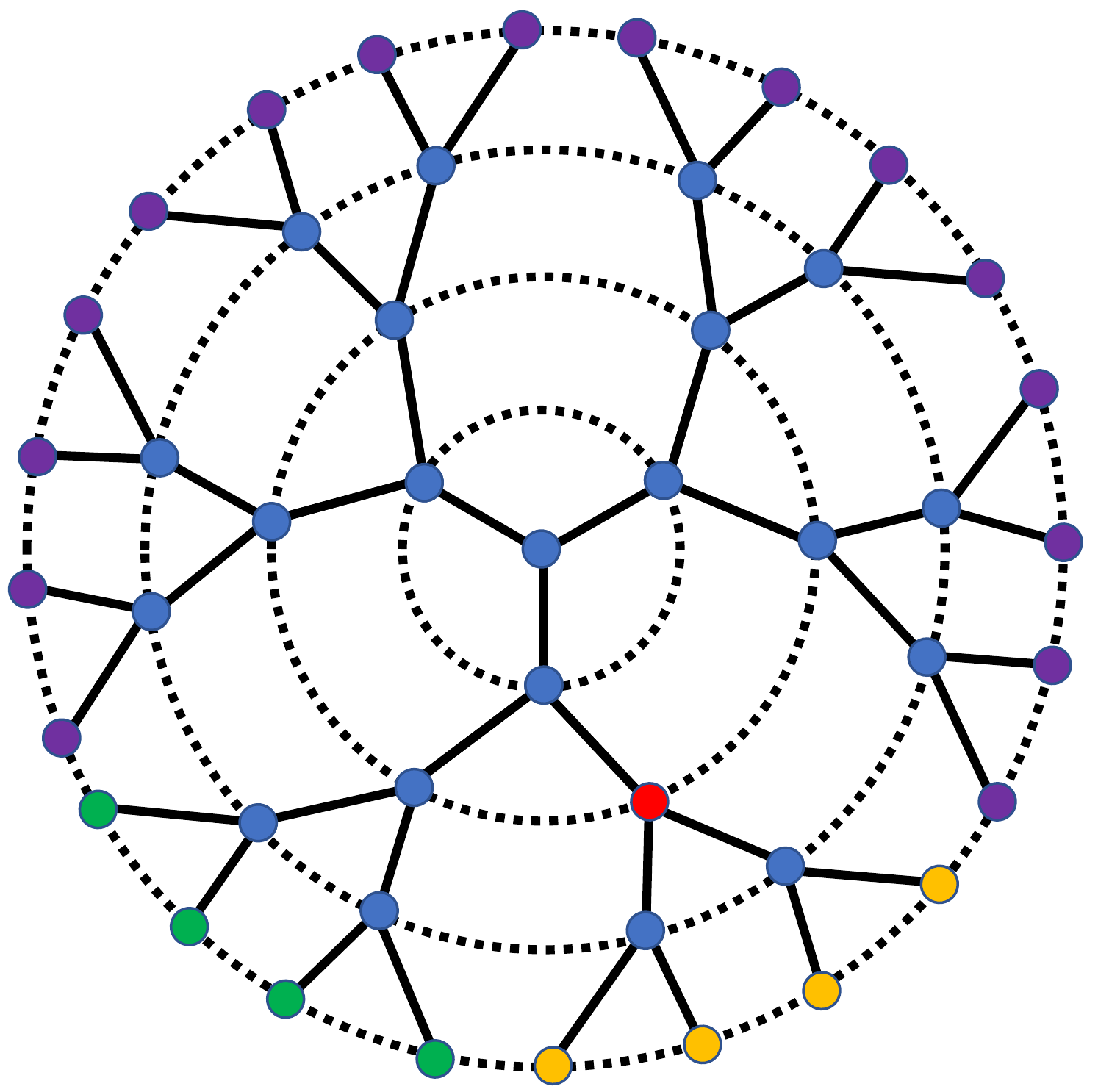}
	\caption{Four generations of the $z = 3$ infinite Bethe lattice emanating from a given center site. The inner shell is at generation $d = 2$, and the outer shell is at generation $d' = 4$. Choosing a given site on the inner shell (colored red), there are three different groups of sites on the outer shell (colored orange, green and purple) that contribute different-length correlations.}
	\label{fig:BL 4 generations}
\end{figure}

\begin{figure}
	\begin{subfigure}[c]{0.49\textwidth}
          \includegraphics[scale=0.9]{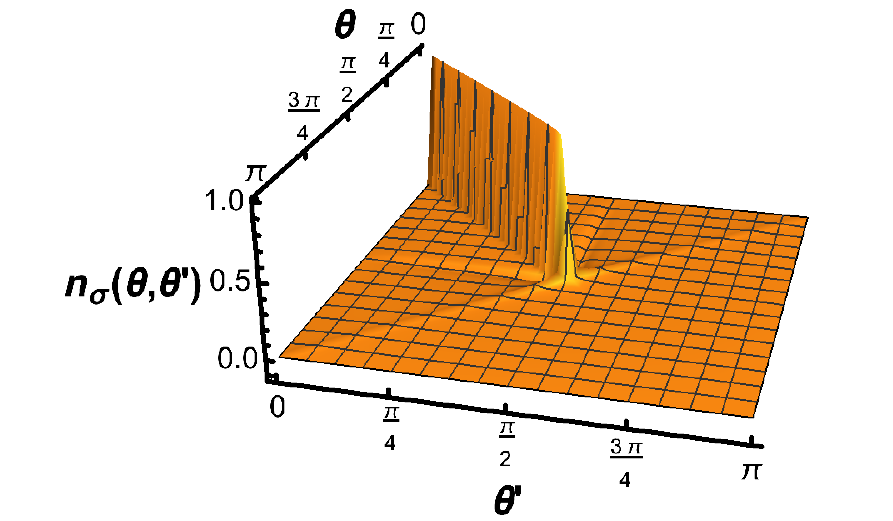}
          \caption{}
	  \label{fig:n_theta_theta_prime_U_0_delmu_0_exact}
        \end{subfigure}\hspace{0.01\textwidth}%
	\begin{subfigure}[]{0.49\textwidth}
          \includegraphics[scale=0.8]{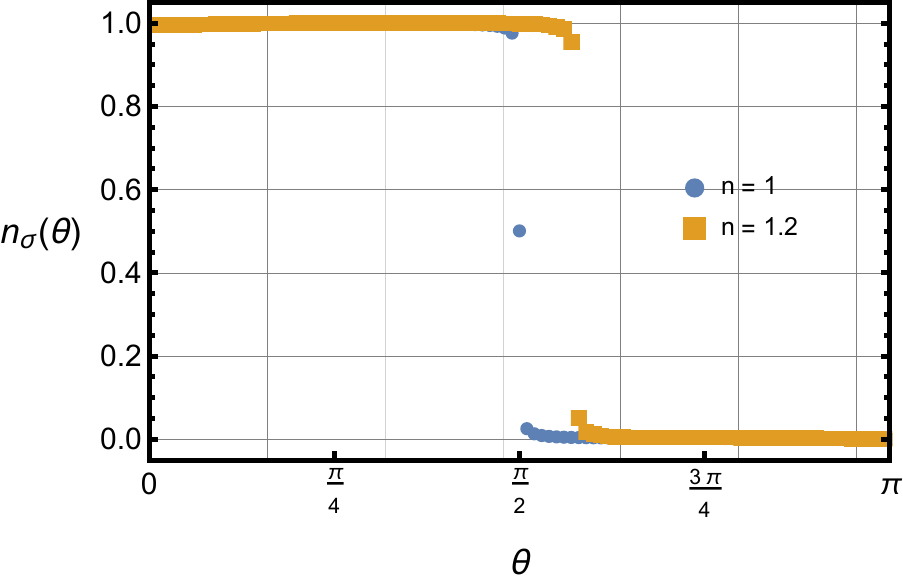}
          \caption{}
          \label{fig:n_theta_U_0_exact}
        \end{subfigure}
	\caption{(a) The occupation function $n_{\s}(\theta, \theta')$ for the half-filled case at $U = 0$. Aside from the expected step-function along the diagonal, there is non-trivial off-diagonal structure. (b) The diagonal component $n_{\s}(\theta)$ for $U = 0$ and densities $\<n_i\> = 1,1.2$, corresponding to values of $\th_F = \pi/2$ and $\th_F \approx 1.81$, respectively. Calculating the occupation from Eq. (\ref{eq:n_theta final}) requires a large distance cutoff $L$, which is chosen here to be (a) $L = 100$ and (b) $L = 200$. This introduces an artificial correlation length, causing the step function to be (slightly) smoothed out.}
	\label{fig:n_2D_and_1D_U_0_exact}
\end{figure}

\begin{figure}[!htbp]
	\begin{subfigure}[c]{0.49\textwidth}
          \includegraphics[scale=0.8]{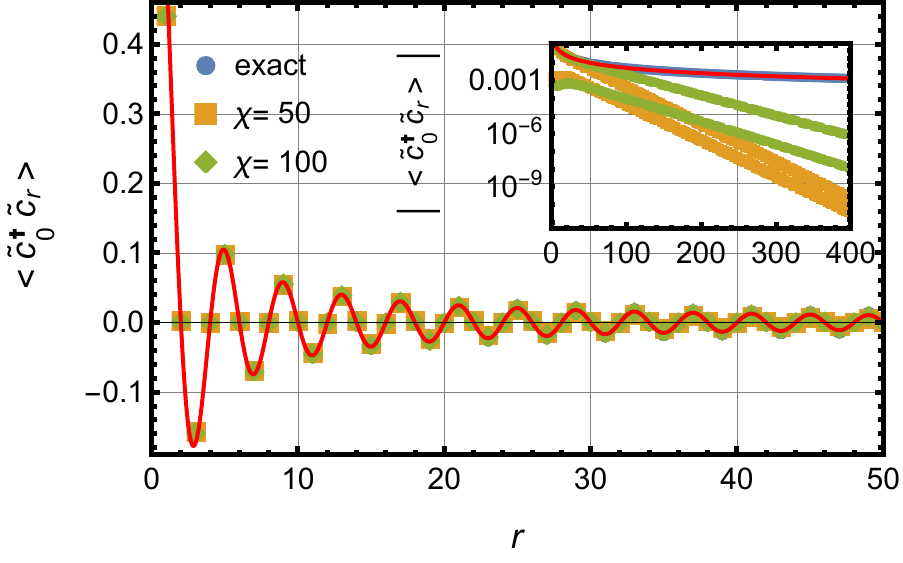}
          \caption{$U = 0, \<n_i\> = 1$.}
          \label{fig:corr function U=0 n = 1}
        \end{subfigure}\hspace{0.01\textwidth}%
	\begin{subfigure}[c]{0.49\textwidth}
          \includegraphics[scale=0.8]{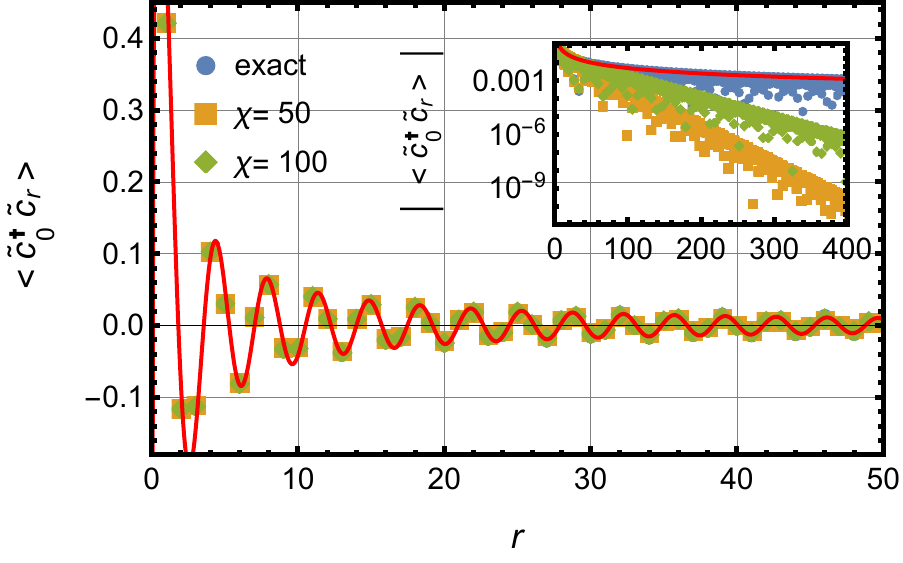}
          \caption{$U = 0, \<n_i\> = 1.2$.}
          \label{fig:corr function U=0 n = 1.2}
        \end{subfigure}
	\begin{subfigure}[c]{0.49\textwidth}
          \includegraphics[scale=0.8]{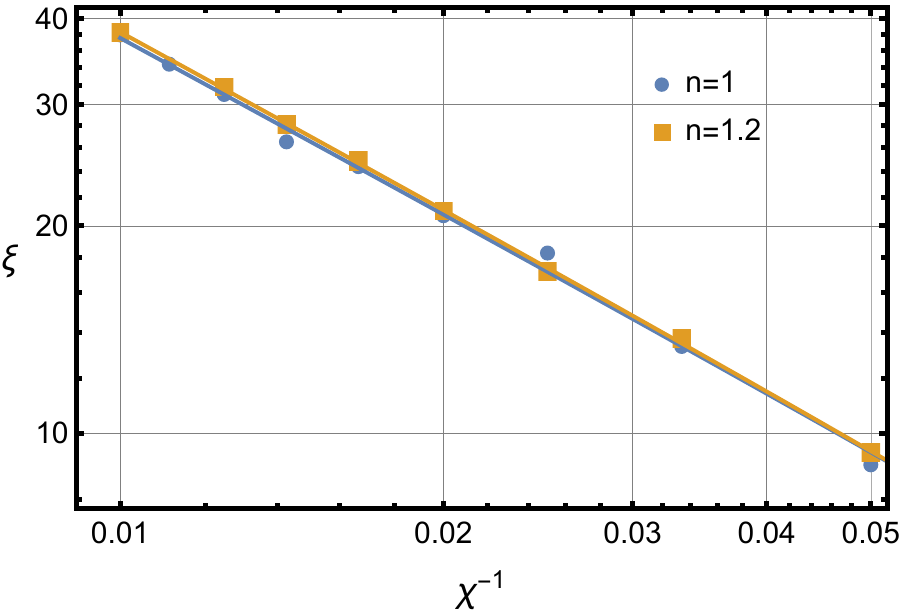}
          \caption{}
          \label{fig:xi vs chi}
        \end{subfigure}
	\caption{The correlation function $\< \td c^{\dag}_{0,\uparrow} \td c_{r,\uparrow}\>$ ($\s = \downarrow$ gives the same) for densities (a) $\<n_i\> = 1$ and (b) $\<n_i\> = 1.2$ for the non-interacting case $U=0$. We show the $\chi = 50,100$ results along with the exact solution of Eq. (\ref{eq:correlation function U=0 exact}). Also shown are the functions $\frac{1}{r} \, \psi_r(\theta_F) \, \psi_0(\theta_F)$ with (a) $\theta_F = \pi/2$ and (b) $\theta_F \approx 1.81$, which are excellent fits beyond a short distance scale, showing Friedel oscillations due to the Fermi surface singularity. The insets show that the $1/r$ decay is fit perfectly over a larger distance by the exact solution, while the finite $\chi$ solutions display an exponential decay at large distances.
		(c) The correlation length extracted from the long distance behavior of $\< \td c^{\dag}_{0,\s} \td c_{r,\s}\>$, plotted versus $1/\chi$. The slope for both densities is $\xi(\chi) \sim \chi^{0.84}$.}
	\label{fig:corr function U=0 real space}
\end{figure}

\begin{figure}
	\includegraphics[scale=0.8]{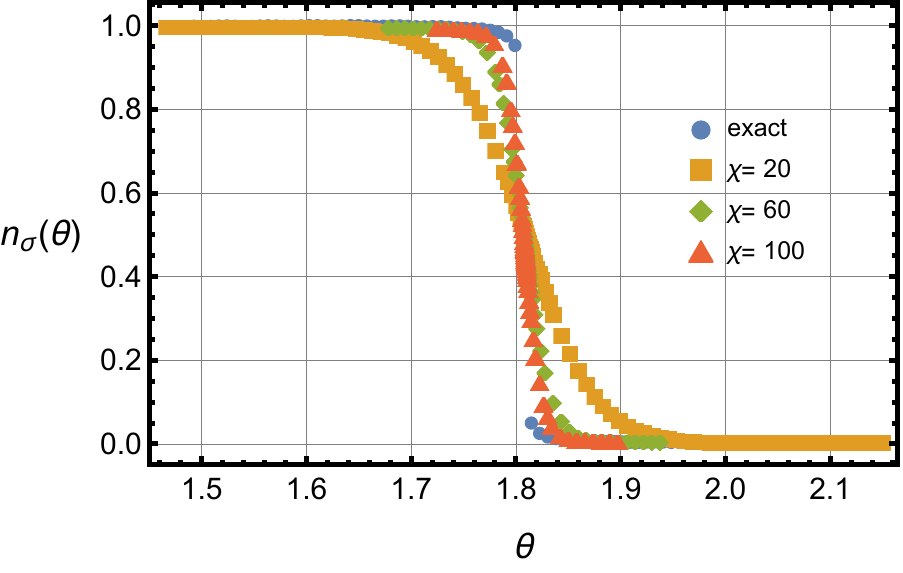}
	\caption{Plots of $n_{\s}(\th)$ near $\th_F$ for $U = 0$ at density $\<n_i\> = 1.2$ and a range of $\chi$. The value of $L$ used here from Eq. (\ref{eq:n_theta final}) is $L = 400$.}
	\label{fig:n_theta near theta_F at U=0 and n=1.2}
\end{figure}

\begin{figure}
	\includegraphics[scale=0.9]{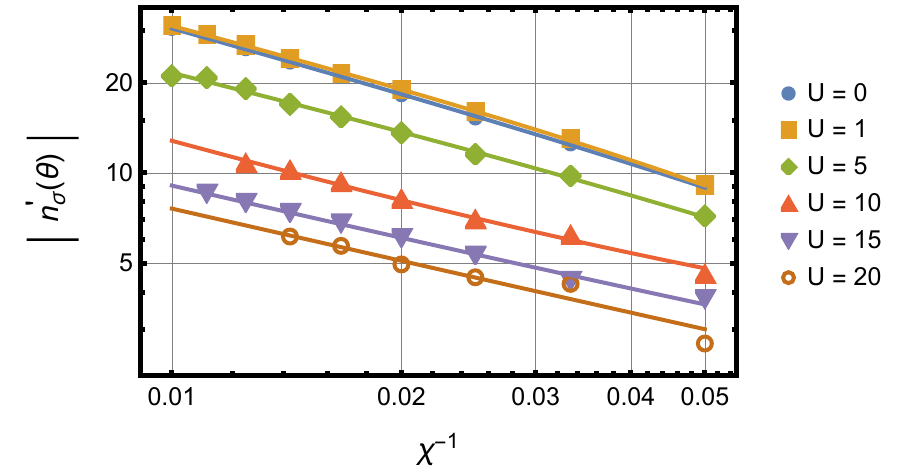}
	\caption{The value of $\abs{n'_{\s}(\th_F)}$ vs $1/\chi$ for various $U$ at density $\<n_i\> = 1.2$. The straight lines on the log-log plot are fit by $\abs{n'_{\s}(\th_F)} \sim \chi ^{\a}$, with $0.5 < \a < 1$.}
	\label{fig:nprime_theta_at_theta_F_vs_chi_all_U_and_n1_2}
\end{figure}

The $\th$-distribution function is calculated from the equal-time correlation functions of symmetric single-particle excitations, $\< \td c^{\dag}_{0,\s} \td c_{l,\s} \>$, where $0$ labels a chosen center site and $l$ labels the generation away from the center site. These can be computed as
\begin{equation}
\begin{split}
\< \td c^{\dag}_{0,\s} \td c_{l,\s}\> 
& = \int\displaylimits_{- \infty}^{0} d\o ~ \mathcal{A}_{0l}(\omega)
\\ & = \int\displaylimits_{-2 \sqrt{z-1}}^{\mu} d\varepsilon \; \sqrt{\frac{z}{2 \pi}} \frac{\psi_l\(\arccos(-\frac{\varepsilon}{2 \sqrt{z-1}})\)}{\sqrt{z^2 - \varepsilon^2}},
\end{split}
\label{eq:correlation function U=0 exact}
\end{equation}
where $\mathcal{A}_{0l}(\omega)$ is the probability of inserting an electron with frequency $\o$ at the center site and observing it at generation $l$ at the same frequency. The occupation function in $\th$-space is defined as
\begin{equation}
n_{\s}(\theta, \theta') \equiv \< \tilde{c}_{\theta,\s}^{\dag} \tilde{c}_{\theta',\s}\>,
\label{eq:occupation function definition}
\end{equation}
where 
\begin{equation}
\td c_{\th, \s} = \lim\limits_{L \to \infty}\sqrt{\frac{\pi}{L+1}} \sum_{d = 0}^{L} \psi_d(\th) \, \td c_{d,\s} 
\label{eq:theta transform c operators}
\end{equation}
are the $\th$-transforms of the symmetric state operators $\td c_{d,\s}$. The key difference with the usual calculation in hypercubic lattices is that $\< \td c^{\dag}_{d,\s} \td c_{d',\s}\>$ is not simply a function of $\abs{d-d'}$, due to the fact that the symmetric states are defined relative to a chosen center site. This is illustrated in Fig. \ref{fig:BL 4 generations}, where we show that $\< \td c^{\dag}_{2,\s} \td c_{4,\s}\>$ contains correlations of length $2,4$ and $6$.
Therefore, $n_{\s}(\theta, \theta')$ is not diagonal in $\th$. One way to understand this is to think of the entire symmetric sector parametrized by $\th$ as the $k = 0$ space on the hypercubic lattice, i.e. the one that is fully symmetric under translations. However, within this sector there is no additional symmetry of the Bethe lattice that requires the total $\th$ to be conserved in a scattering process, and therefore $n_{\s}(\theta, \theta')$ is not diagonal \cite{Eckstein_thanks}. Carefully inserting Eq. (\ref{eq:theta transform c operators}) into Eq. (\ref{eq:occupation function definition}) and rewriting the result in terms of $\< \td c^{\dag}_{0,\s} \td c_{l,\s}\>$ gives
\begin{equation}
\begin{split}
& n_{\s}(\theta, \theta') = 
\lim\limits_{L \to \infty}\frac{\pi}{L+1} \sum_{d,d' = 0}^{L} \psi_d(\th) \psi_{d'}(\th') \, 
\frac{z-2}{\sqrt{z (z-1)}}
\\ &
\(\frac{\sqrt{z}(z-2)}{(z-1)^{3/2}}\)^{\delta_{\min(d,d'),0}} 
\sum_{r=0}^{\min(d,d')} \(\frac{z-1}{z-2}\)^{\delta_{r,0} + \delta_{r,\min(d,d')}} 
\\ & 
\sqrt{\frac{z}{z-1}}^{\delta_{d,d'} \delta_{r,0} - \delta_{d,0} \delta_{d',0}} 
\, \< \td c^{\dag}_{0,\s} \td c_{\abs{d-d'}+2r,\s} \>. 
\end{split}
\label{eq:n_theta final}
\end{equation}
Details of the derivation of Eq. (\ref{eq:n_theta final}) are in Appendix \ref{sec:occupation function}. We plot the exact $n_{\s}(\theta, \theta')$ for $U = 0$ at half-filling in Fig. \ref{fig:n_theta_theta_prime_U_0_delmu_0_exact}.
\\
\indent
In this work, we focus exclusively on the diagonal component of the occupation function, $n_{\s}(\theta) \equiv n_{\s}(\th, \th)$. This tells us the occupation of excitations that preserve total $\th$ when scattering with each other. We leave the detailed study of the full occupation function $n_{\s}(\th, \th')$ for future work. We plot the exact $n_{\s}(\theta)$ in Fig. \ref{fig:n_theta_U_0_exact} for $U = 0$ and densities $\<n_i\> = 1, 1.2$. We can see the expected behavior of $n_{\s} = \Theta(\th_F - \th)$, where $\Theta(x)$ is the Heaviside step function and $\th_F$ is the $\th$-analog of the ``Fermi momentum" (the step function in Fig. \ref{fig:n_theta_U_0_exact} is slightly smoothed out due to a finite $L$ in Eq. (\ref{eq:n_theta final})).
\\
\indent 
In order to compute the correlation function $\< \td c^{\dag}_{0,\s} \td c_{l,\s}\>$ from the VUTS numerical solution, 
we use the fact that even in the interacting ground state, $\< c^{\dag}_{i,\s} c_{j,\s} \>$ is still only a function of the distance $\abs{i-j}$ (and $\s$). We can then compute $\< c^{\dag}_{0,\s} c_{l,\s} \>$ for an arbitrary branch to write
\begin{equation}
\< \td c^{\dag}_{0,\s} \td c_{l,\s}\>
= \sqrt{z^{1-\delta_{l,0}} (z-1)^{l-1 + \delta_{l,0}}} \< c^{\dag}_{0,\s} c_{l,\s}\>. 
\label{eq:correlation function symmetric states}
\end{equation}
Note that the difference here from the previous considerations of this paragraph is that one of the reference points has been set to the center site. We plot $\< \td c^{\dag}_{0,\s} \td c_{l,\s}\>$ measured with VUTS in Fig. \ref{fig:corr function U=0 real space} for $U = 0$ and densities $\< n_i \> = 1, 1.2$, along with the exact results. 
The finite $\chi$ results are close to the exact ones, but, as expected, the correlation function at large enough distances decays exponentially with a finite correlation length $\xi(\chi)$. We can measure $\xi(\chi)$ by fitting $\< \td c^{\dag}_{0,\s} \td c_{l,\s}\>$ to an exponential at large $l$. The results are shown in Fig. \ref{fig:xi vs chi}, where the polynomial fit gives $\xi(\chi) \sim \chi^{0.84}$ for both densities.
\\
\indent 
When calculating $n_{\s}(\th)$ from Eq. (\ref{eq:n_theta final}), in practice we must choose a finite value of $L$. As long as we take $L$ large enough, the correlation length $\xi(\chi)$ will act as the long-distance cutoff and the value of $L$ will not have any effect. Our results are obtained using bond dimensions up to $\chi = 100$, for which the induced correlation length is $\xi(\chi) \lesssim 40$. We find that a value of $L \sim 400$ is large enough for all $\chi$ we study. The finite $\xi(\chi)$ smoothes out the step function in $n_\s(\th)$, so we estimate $\th_F$ from the location of the maximum of $\abs{n'_{\s}(\th)}$. In Fig. \ref{fig:n_theta near theta_F at U=0 and n=1.2} we show $n_{\s}(\th)$ for $\<n_i\> = 1.2$ near $\th_F$. 
The finite slope at $\th_F$ diverges as a power law in $\chi$, as we show in Fig. \ref{fig:nprime_theta_at_theta_F_vs_chi_all_U_and_n1_2}. 
This indicates that $\xi(\chi)$ is the only low-energy scale in the problem, and the state is truly gapless in the $\chi \to \infty$ limit.
\\
\indent
In summary, in this section we show how to compute the diagonal part of the occupation function in the non-interacting case, using VUTS and 
a careful extrapolation in the bond dimension, and achieve excellent agreement with the analytical result.

\subsection{Quasiparticles in the interacting system}

\begin{figure}
	\begin{subfigure}[]{0.49\textwidth}
          \includegraphics[scale=0.8]{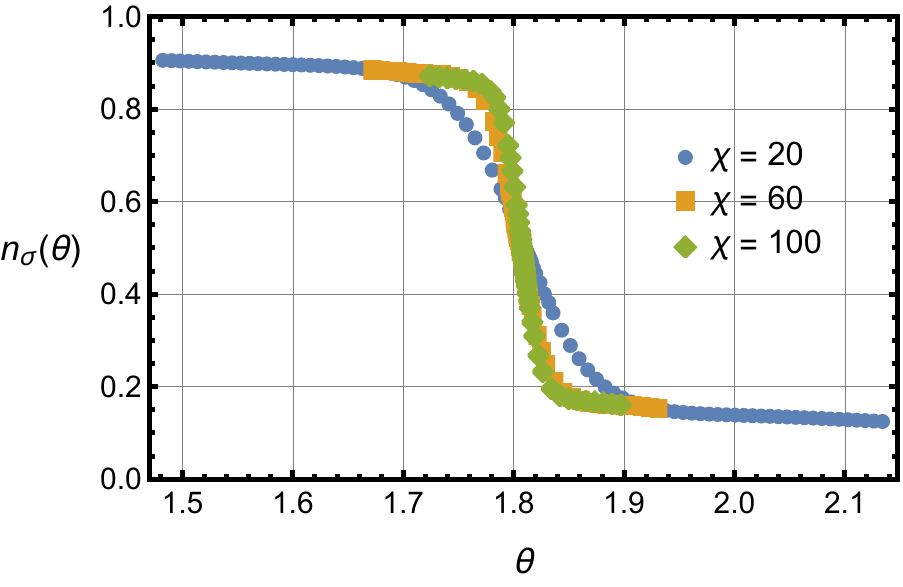}
          \caption{$\<n_i\> = 1.2, U = 5$.}
        \end{subfigure}\hspace{0.01\textwidth}%
	\begin{subfigure}[]{0.49\textwidth}
          \includegraphics[scale=0.8]{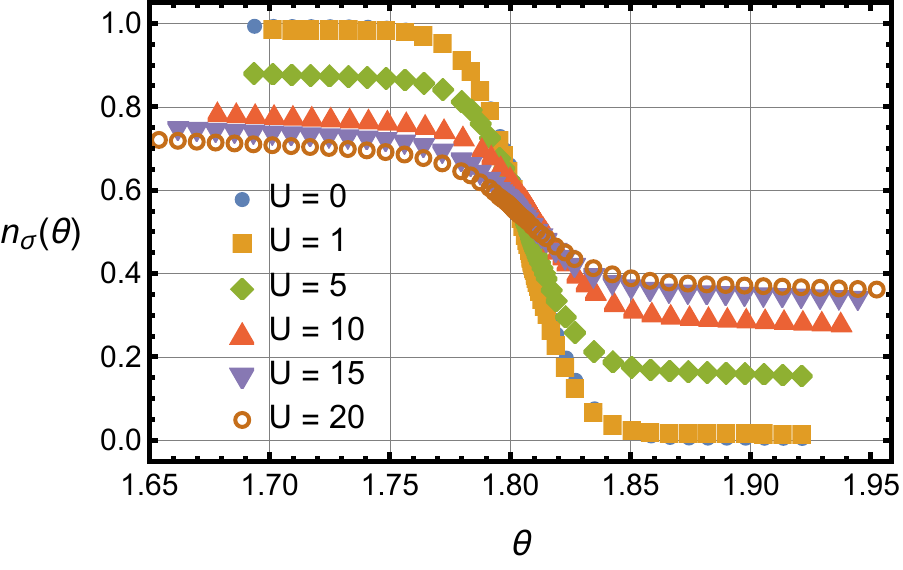}
          \caption{$\<n_i\> = 1.2, \chi = 70$.}
        \end{subfigure}
	\caption{Plots of $n_{\s}(\th)$ near $\th_F$ at density $\<n_i\> = 1.2$, showing the dependence on (a) $\chi$ for fixed $U = 5$, and on (b) $U$ for fixed $\chi = 70$. The value of $L$ from Eq. (\ref{eq:n_theta final}) is $L = 400$.}
	\label{fig:n_theta near theta_F all U and n=1.2}
\end{figure}

\begin{figure}
	\includegraphics[scale=0.9]{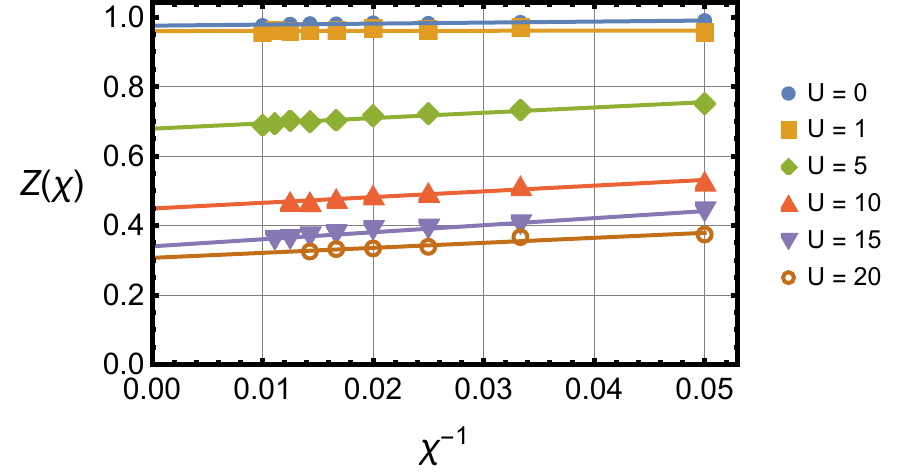}
	\caption{$Z(\chi)$ at density $\<n_i\> = 1.2$, shown with linear fits.}
	\label{fig:Z vs chi for all U at n=1.2}
\end{figure}

\begin{figure}
	\begin{subfigure}[]{0.49\textwidth}
    	\includegraphics[scale=0.8]{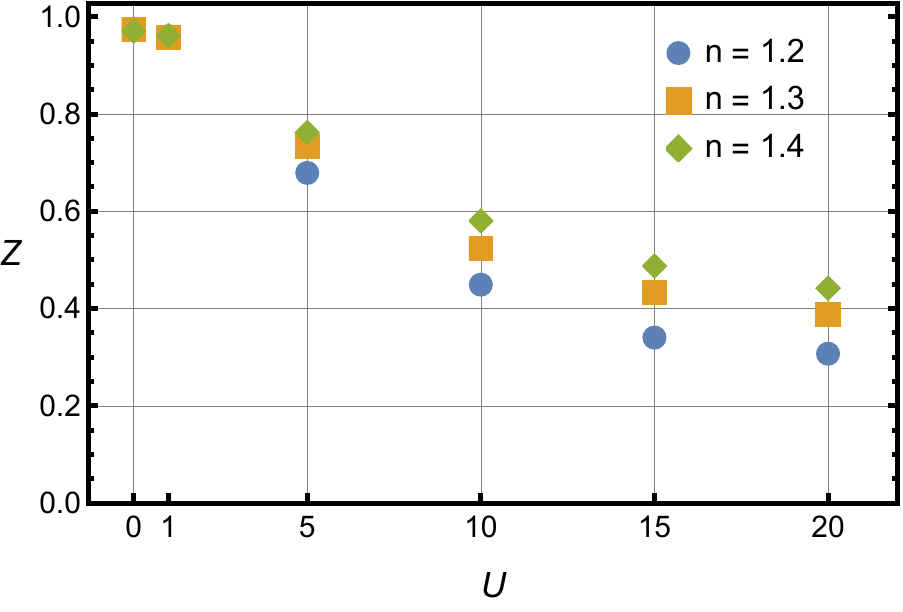}
        \caption{}
	    \label{fig:Z vs U for all n}
        \end{subfigure}
	\begin{subfigure}[]{0.49\textwidth}
    	\includegraphics[scale=0.8]{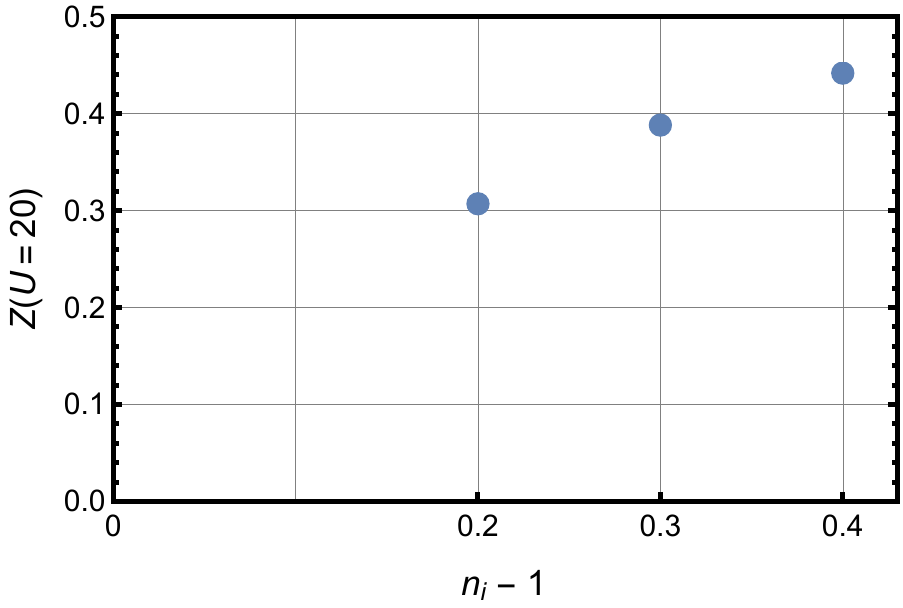}
    	\caption{}
	    \label{fig:Z at U =20 vs n-1}
        \end{subfigure}
    	\caption{(a) The extrapolated values of $Z$ at densities $\<n_i\> = 1.2,1.3,1.4$, plotted as a function of $U$. The first two blue points are covered by the orange ones. The error bars, which are the discrepancy in the extrapolation with and without the last data point, are smaller than the data points. We can see that the curves are close to saturation at $U = 20$. (b) The values of $Z$ at $U = 20$, which seem close to the saturated values, as a function of $\<n_i\> - 1$.}
\end{figure}

Now we turn on interactions. In Fig. \ref{fig:n_theta near theta_F all U and n=1.2} we plot $n_{\s}(\th)$ near $\th_F$ for $\<n_i\> = 1.2$ and $U = 5$ at various $\chi$ as well as for various $U$ at $\chi = 70$.
The occupation shows a form similar to that of the free case, albeit with a reduced size of the step at $\th_F$. The value of the slope at $\th_F$ diverges for all $U$, as we show in Fig. \ref{fig:nprime_theta_at_theta_F_vs_chi_all_U_and_n1_2}. The interacting system is therefore also gapless in the $\chi \to \infty$ limit, as expected.
\\
\indent 
The quasiparticle weight, $Z$, of the symmetric state excitations is defined as 
\begin{equation}
Z \equiv \(\lim\limits_{\th \to \th_F^-} - \lim\limits_{\th \to \th_F^+} \) n_{\s}(\th).
\label{eq:Z definition}
\end{equation}
For a Fermi liquid $n_{\s}(\th)$ has a step at $\th_F$ and $Z > 0$, while for a Luttinger liquid the occupation function has a higher order non-analyticity that scales as $n_{\s}(\th - \th_F) \sim \abs{\th - \th_F}^{\g} \sign(\th - \th_F)$ for some $\g < 1$ and therefore $Z = 0$. Our goal is to find the true thermodynamic value of $Z$ to distinguish between these two scenarios. Of course, for a finite $\chi$ Eq. (\ref{eq:Z definition}) will always give zero. However, we can define a quantity $Z(\chi)$ whose limit will give $Z$ in the $\chi \to \infty$ limit. We define this as
\begin{equation}
Z(\chi) \equiv n_{\s}\(\th_F - \frac{\pi}{2 \, \xi (\chi)} \) - n_{\s}\(\th_F + \frac{\pi}{2 \, \xi (\chi)} \),
\label{eq:Z finite chi definition}
\end{equation}
which satisfies the desired property because $\xi(\chi)\rightarrow \infty$ with increasing $\chi$. We choose a spacing of $\Delta \th = \pi/\xi(\chi)$ around $\th_F$ because that is roughly the resolution one expects from a finite correlation length, and therefore the convergence in $1/\chi$ should be fastest. We plot $Z(\chi)$ vs $\chi$ for $\<n_i\> = 1.2$ and a range of $U$ in Fig. \ref{fig:Z vs chi for all U at n=1.2}. 
The results show that the extrapolated $Z$ is (a) very close to the expected value of $Z = 1$ for the free theory and (b) finite for all $U$ we study. We plot $Z$ as a function of $U$ in Fig. \ref{fig:Z vs U for all n}, where we can see that it decreases as a function of $U$, but seems to saturate to a finite value. The saturation value is an increasing function of doping $\<n_i\> - 1$, as 
illustrated by Fig.~\ref{fig:Z at U =20 vs n-1} where we plot the value of $Z(U=20)$ as a function 
of doping. 

We also address the question of Luttinger's theorem for $n_{\s}(\th)$. We check that $\theta_F$ is independent of $U$ (the dependence on $\chi$ is negligible) for various densities  in the range $\<n_i\> \in (1.1,1.4)$. From this we conclude that Luttinger's theorem holds for all values of the density and interaction strength. 
\\
\indent
The decrease of $Z$ with increasing $U$ as well as the saturation at large $U$ to a value which increases with doping are both qualitatively consistent with results established in the $z=\infty$ (DMFT) limit and with slave boson approaches (for general lattices) ~\cite{RevModPhys.68.13,4bosons}.
A distinctive aspect of these theories, however, is that the effective mass of 
quasiparticles is related to $Z$ by $m^*/m=1/Z$. On the technical level, this is due to the locality of the self-energy, while physically this reflects the inability of these approaches to capture the feedback of short-range order and collective modes 
into the physics of quasiparticles. 
In contrast, in the present case, since the connectivity is kept finite, we would expect this feedback to be present. 
It is therefore an outstanding question for future work to explore whether the dispersion of quasiparticles is 
renormalized in a  different manner than $Z$ itself, and in particular whether it is affected by 
short-range antiferromagnetic correlations at low doping level. This is left for future work since it requires 
an extension of our algorithm to the study of excited states.

\section{Discussion}
\label{sec:discussion}

In summary, we have introduced a new numerical algorithm, 
(fermionic) VUTS, to study quantum (fermionic) models on the Bethe lattice. 
We apply it to the Hubbard model for coordination number $z = 3$, allowing for a two-site unit cell, obtain the $T = 0$ phase diagram and study the doping-induced Mott transition.   
We find an antiferromagnetic insulating phase at half filling and a paramagnetic metallic phase for the doped system, which are separated by a first-order insulator to metal phase transition. 
The model displays phase separation at low doping, with a range of forbidden densities. These conclusions were reached by allowing for a two-site unit cell. We cannot exclude that phases with more complex charge or 
magnetic ordering exist when allowing for a larger unit cell, which we leave as an open question for future work. By studying the diagonal component of the occupation function for momenta of the symmetric single-particle sector, we find that the quasiparticle weight is non-zero, 
consistent with the existence of a Fermi liquid ground state for fermions on the Bethe lattice. We find that this Fermi liquid state obeys Luttinger's theorem.

An interesting direction in which to extend this work would be to further characterize this Fermi liquid state. 
One would like to know, for example, what happens near $\th_F$ to the off-diagonal $n_\s(\th,\th')$ when interactions are turned on. 
It is also interesting to look at some of the other symmetry sectors, say the ones that leave each of the $z$ sub-trees connected to the center site invariant, 
and see whether they have quasiparticles and how the quasiparticle weight depends on the sector.
These questions are the Bethe lattice version of the important physical question of `momentum dependence' of quasiparticle properties on the Fermi surface of hypercubic lattices.

In the one-dimensional VUMPS algorithm, it has been shown that low-lying excitations above the ground-state can be accurately computed~\cite{PhysRevB.97.235155}. 
An obvious question is how to extend these ideas to VUTS (this is another potential advantage of this method over imaginary time evolution).
Extension of the algorithm to excited states would allow to characterize the effective mass (dispersion) of the quasiparticles, paving the road for a study of 
how short-range (e.g. antiferromagnetic) correlations affect quasiparticle properties. This is a very important question for the physics of strongly correlated electron systems. 
A breakdown of the $m^*/m=1/Z$ relation would signal that this feedback is indeed present, in contrast to the infinite connectivity limit. 
Finally, studying the energy dependence of the quasiparticle lifetime, as well as the interactions between quasiparticles would be a comprehensive study of Landau 
Fermi liquid theory on the Bethe lattice. 

It would also be interesting to look at the entanglement structure of the Fermi liquid state, as compared to a Luttinger liquid. Since the entanglement spectrum is easily obtained on the one-dimensional and Bethe lattices from the singular value decomposition of a single bond tensor, this question could in principle be easily answered. 

We also propose that the finite $z$ Bethe lattice can be used as a computationally tractable platform for the study of how quasiparticles can be destroyed 
and Fermi liquid behaviour breaks down when considering other fermionic hamiltonians on this lattice. 
One route to explore this, which connects to the feedback of long-wavelength collective modes or short-range spatial correlations on quasiparticle properties, is to study the vicinity of a quantum critical point. 
Another route is to study microscopic models that are tailor-engineered to have incoherent excitations, such as the one of Ref.~\cite{PhysRevB.86.045128}, or the multi-channel Kondo lattice.     

Studying the interplay between frustration and strong correlations is another promising direction for future research. 
Introducing a frustrating next-nearest-neighbor hopping term has been found with DMFT to yield an interaction-driven metal-insulator transition on the infinite $z$ Bethe lattice~\cite{rozenberg1994,RevModPhys.68.13}. An interesting question is whether this transition can also be found at finite $z$. 
The VUTS algorithm could also be extended to other lattices with a tree-like structure, such as the Husimi cactus. 
The study of spin models on such lattices have revealed spin-liquid ground-states\cite{Chandra_1994,PhysRevB.93.075154} (see also Ref.~\cite{Udagawa_2019} for considerations on the spin-ice model), 
opening the question of how these models behave upon doping. 

Finally, increasing the temperature to a non-zero value is another interesting direction. Intuitively, some finite-temperature properties may be less sensitive to the differences between tree lattices and two- or three-dimensional hypercubic lattices. This could be done using the purification method \cite{PhysRevLett.93.207205,PhysRevLett.93.207204,PhysRevB.72.220401} which has been formulated on the Bethe lattice in Ref. \cite{PhysRevB.100.125121}.

\section*{Acknowledgments}

We thank Julien Agier, Martin Claassen, Michel Ferrero, Gabriel Kotliar, Chris Laumann, Sung-Sik Lee, Roderich Moessner, Marcelo Rozenberg, Steve White and in particular Martin Eckstein and Riccardo Rossi for useful discussions. We thank Ruben Verresen for comments on the manuscript. All tensor network calculations and the VUTS code were implemented with the ITensor Library (C++ version 3.1) \cite{itensor}. The Flatiron Institute is a division of the Simons Foundation.


\bibliographystyle{apsrev4-1}
\bibliography{references}

\appendix
\widetext

\setcounter{page}{1}
\setcounter{figure}{0}
\setcounter{subfigure}{0}
\setcounter{table}{0}
\setcounter{equation}{0}


\renewcommand\thefigure{A\arabic{figure}}
\renewcommand\thetable{A\arabic{table}}
\renewcommand\theequation{A\arabic{equation}}

\section{Details of the variational uniform tree states algorithm}
\label{sec:VUTS details}

\subsection{Summing the Hamiltonian terms}
\label{subsec:summing Hamiltonian terms}

\begin{figure}
	\begin{subfigure}[c]{0.49\textwidth}
	    \includegraphics[scale=0.25]{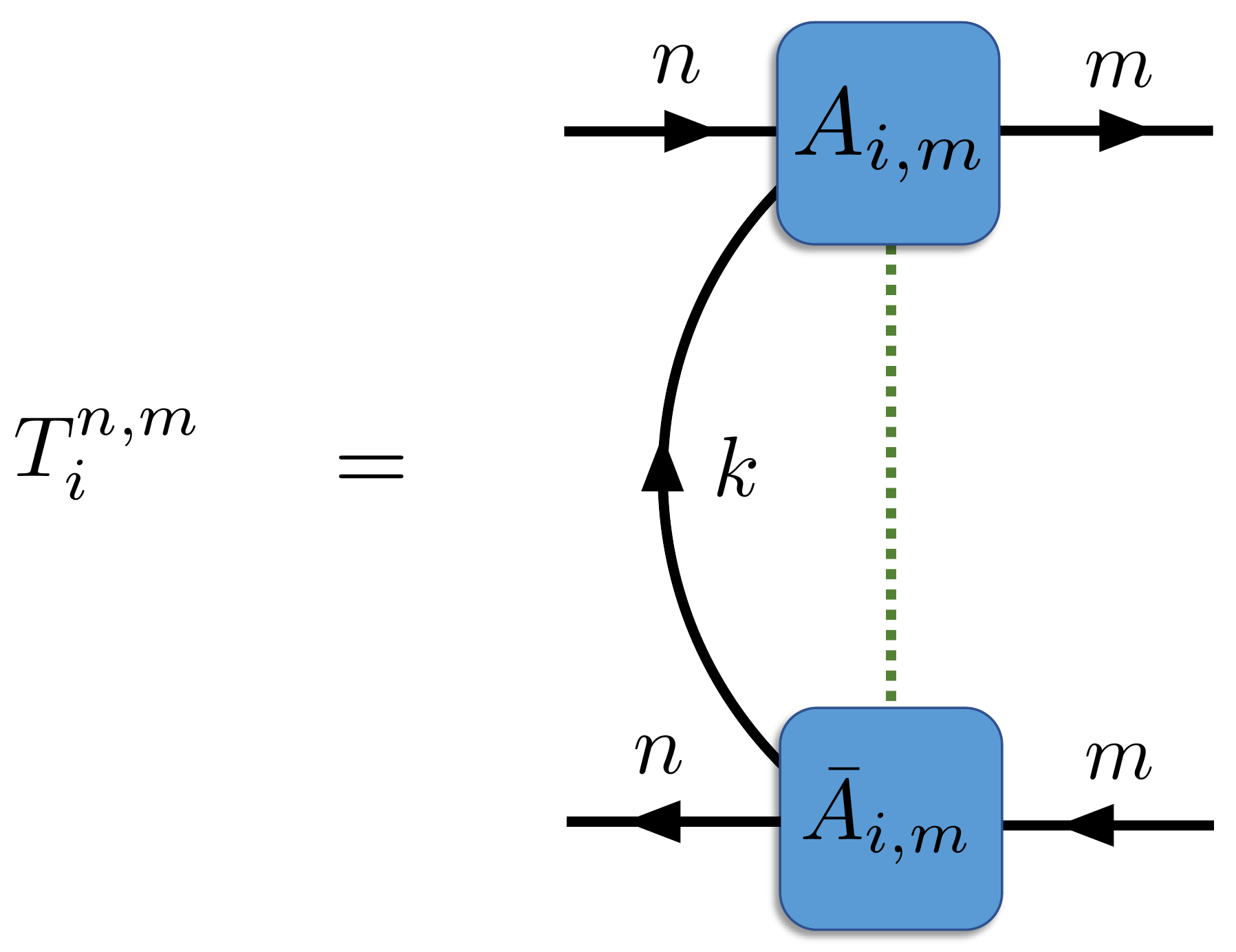}
	    \caption{}
	    \label{fig:transfer matrix single}
        \end{subfigure}\hspace{0.01\textwidth}%
	\begin{subfigure}[c]{0.49\textwidth}
        \includegraphics[scale=0.25]{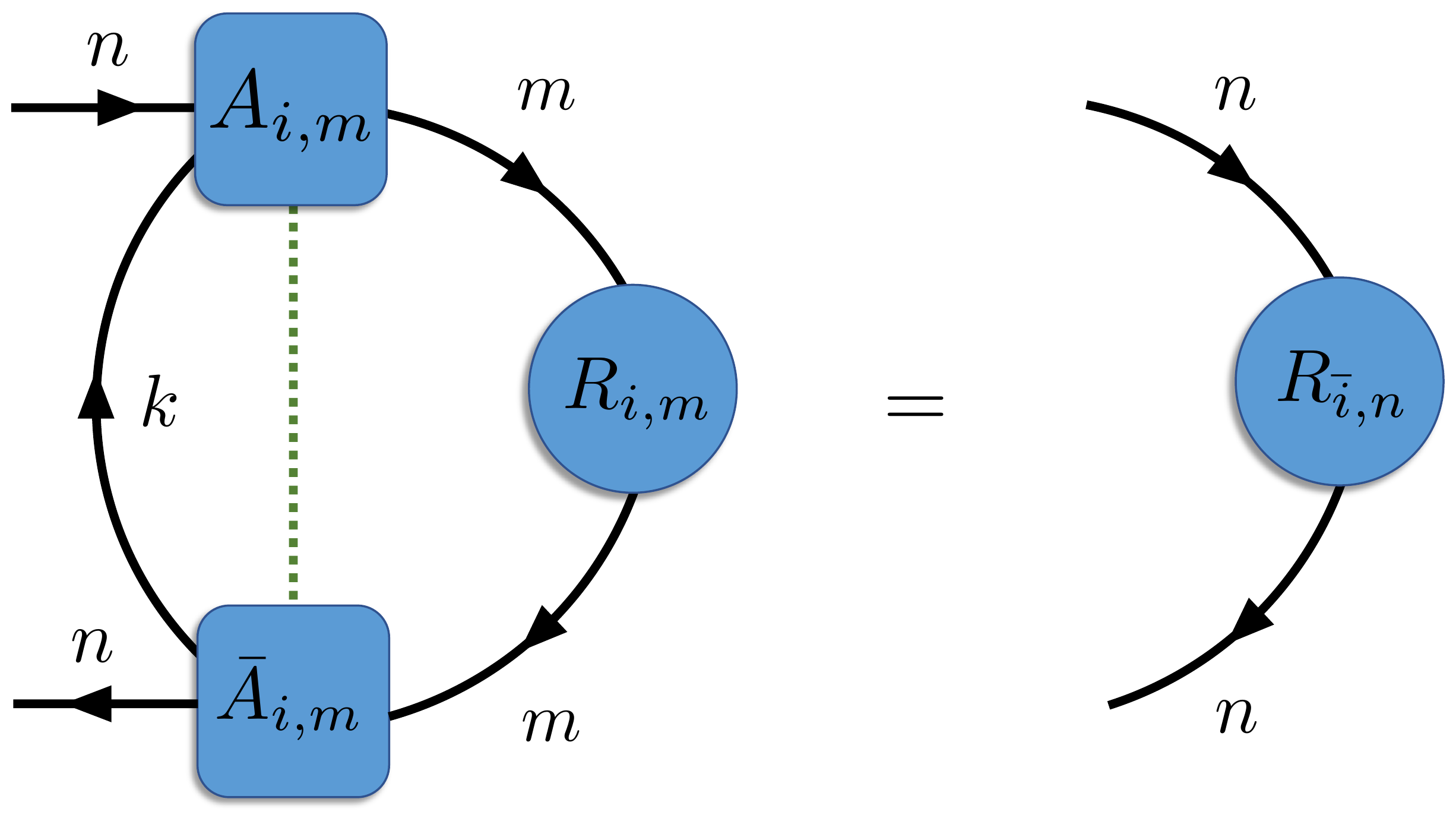}
        \caption{}
	    \label{fig:reduced density matrices}
        \end{subfigure}\hspace{0.01\textwidth}%
    \caption{(a) Transfer matrix $T_i^{n,m}$ from Eq.     (\ref{eq:transfer matrix one site}). (b) Diagrammatic equations for the right eigenvectors $R_{i,m}$ of the transfer matrices, which represent the reduced density matrices for bipartitions of the Bethe lattice (and are also shown in Eq. (\ref{eq:reduced density matrices})). As a reminder, the bonds labelled $k,m,n$ have link indices $l_k,l_m,l_n$ respectively, and we leave of the prime levels for simplicity.}        
    \label{fig:transfer matrices and reduced density matrices}
\end{figure}

Here we explain how we compute projected Hamiltonians for subtrees of the Bethe lattice, using infinite summations like those shown in Fig. \ref{fig:H_{1,2} series}. To help us with the computation, we start by defining the \emph{transfer matrices}, which are operators that act on the virtual degrees of freedom of the network. We build the transfer matrices out of the $A_{i,m}$ tensors, the tensors of the TTN that are in the orthonormal gauge. They are defined as
\begin{equation}
T_{i}^{n,m} 
= \sum_{s_i,l_k} \bar{A}^{s_i,l'_m,l'_n,l_k}_{i,m} A^{s_i,l_m,l_n,l_k}_{i,m},
\label{eq:transfer matrix one site}
\end{equation}
where $n, m = 0, 1, 2$ and $n\ne m$. The diagrammatic version of Eq. (\ref{eq:transfer matrix one site}) is shown in Fig. \ref{fig:transfer matrix single}. 
Using the transfer matrix we can rewrite the gauge conditions of Eq. (\ref{eq:canonical gauge}) (Fig. \ref{fig:canonical gauge}) as 
\begin{equation}
(\mathbb{1}_{\bar i,n}| \; T_{i}^{n,m} = (\mathbb{1}_{i,m}|,
\end{equation}
where we use the vectorized notation such that $(\mathbb{1}_{i,m}|$ is a length $\chi^2$ vector that is isomorphically equivalent to the $\chi \times \chi$ matrix $\mathbb{1}_{i,m}$, and the transfer matrix $T_{i}^{n,m}$ is treated as a $\chi^2 \times \chi^2$ matrix mapping from the bra and ket links of bond $n$ to the bra and ket links of bond $m$.
In addition, we make use of the right eigenvectors of the transfer matrices, $R_{i,m}$. These are defined by the set of equations:
\begin{equation}
\displaystyle\sum_{s_i,l_k,l_m,l'_m} \bar{A}^{s_i,l'_m,l'_n,l_k}_{i,m} \, R^{l'_m,l_m}_{i,m}  \, A^{s_i,l_m,l_n,l_k}_{i,m} = R^{l'_n,l_n}_{\bar i,n}, 
\label{eq:reduced density matrices}
\end{equation}
or in vectorized notation in terms of the transfer matrices:
\begin{equation}
T_{i}^{n,m} \; |R_{i,m}) = |R_{\bar i,n}).
\label{eq:reduced density matrices using transfer matrix}
\end{equation}
The matrices $R_{i,m}$ are the reduced density matrices of a bipartition of the state. The diagrammatic versions of Eqs. (\ref{eq:reduced density matrices}) and (\ref{eq:reduced density matrices using transfer matrix}) is shown in Fig. \ref{fig:reduced density matrices}. 
\\
\indent
Using vectorized notation, the environment matrices $H_{i,m}$ are calculated from the geometric series
\begin{equation}
(\mathbf{H}_i|
=
\[
(\mathbf{h}_i|
+
(\mathbf{h}_{\bar i}|
\mathbf{T}_i
\]
\displaystyle\sum_{k = 0}^{\infty}
\(
\mathbf{T}_{\bar i}
\mathbf{T}_i
\)^k
\label{eq:environment tensor geometric series}
\end{equation}
where $(\mathbf{H}_i|$ and $(\mathbf{h}_i|$ are vectors of vectorized matrices:
\begin{equation}
(\mathbf{H}_i| =
\begin{pmatrix}
(H_{i,0}| &
(H_{i,1}| &
(H_{i,2}|
\end{pmatrix}
,\hspace{0.025\textwidth}%
(\mathbf{h}_i| =
\begin{pmatrix}
(h_{i,0}| &
(h_{i,1}| &
(h_{i,2}|
\end{pmatrix}
\end{equation}
and $\mathbf{T}_i$ is a matrix of transfer matrices:
\begin{equation}
\mathbf{T}_i =
\begin{pmatrix}
0         & T_i^{0,1} & T_i^{0,2} \\
T_i^{1,0} & 0         & T_i^{1,2} \\
T_i^{2,0} & T_i^{2,1} & 0
\end{pmatrix}.
\end{equation}
Here, the $(h_{i,m}|$ are vectorized version of the tensors $h_{i,m}$, which are the projections of a local Hamiltonian term $h_{i,j}$ into the orthonormal basis of $A_{i,m}$, and are defined diagrammatically in Fig. \ref{fig:h_i_m}. We remind readers that the transfer matrices of the $z=3$ Bethe lattice are $T_i^{m,n}$ for $m, n = 0, 1, 2$ and $m \ne n$. We start by analytically summing Eq. (\ref{eq:environment tensor geometric series}), using the fact that it is of the form of a geometric series (in other words, using that $\sum_{k=0}^{\infty} x^k = (1-x)^{-1}$, where $x = \mathbf{T}_{\bar i} \mathbf{T}_i$ is the product of the two matrices of transfer matrices). However, we also have subtract away the infinite number of local energy contributions (one from each term in the series) which cause the series to diverges. This results in the following linear set of equations for $(H_{i,m}|$:
\begin{equation}
\begin{split}
(H_{i,m}| 
-
\displaystyle\sum_{m',m''} (H_{i,m''}|  \, T_{\bar{i}}^{m'',m'} T_{i}^{m',m}
\(
\mathbf{\mathbbm{1}}
-
|R_{i,m}) (\mathbb{1}_{i,m}|
\)
=
\(
(h_{i,m}|
+
\displaystyle\sum_{m'} (h_{\bar i,m'}|  \, T_{i}^{m',m}
\)
\(
\mathbf{\mathbbm{1}}
-
| R_{i,m}) (\mathbb{1}_{i,m}|
\)
.
\end{split}
\label{eq:environment tensor}
\end{equation}
The sums of $m'$ and $m''$ run over $0,1,2$ (for each direction in the $z=3$ Bethe lattice), but we remind readers that $T_i^{m,n}$ only exists for $m\ne n$.
The terms proportional to $(\mathbb{1}_{i,m}|$ on the left hand side of Eq. (\ref{eq:environment tensor}) act to project out the dominant eigenspace of the transfer matrix that contains the infinite energy contribution, and lead to a convergent geometric series. Congruently, the terms proportional to $(\mathbb{1}_{i,m}|$ on the right hand side of Eq. (\ref{eq:environment tensor}) zero out the on-site energies of the unit cell. Eq. (\ref{eq:environment tensor}) is a generalization of the result derived in Appendix D of Ref. \cite{PhysRevB.97.045145} to the $z=3$ Bethe lattice with a 2-site unit cell. We now can think of the solution to Eq. (\ref{eq:environment tensor}) as defining $(H_{i,m}|$, rather than defining it via the geometric series of Eq. (\ref{eq:environment tensor geometric series}). Applying the vectorized density matrices $|R_{i,m})$ to Eq. (\ref{eq:environment tensor}), we see that $(H_{i,m}|$ satisfies $(H_{i,m}|R_{i,m}) = 0$, so the infinite-energy shift can be seen as shifting the total energy of the environment to zero. We can see that the only unknowns in Eq. (\ref{eq:environment tensor}) are the three environment matrices $(H_{i,m}|$. Also note that Eq. (\ref{eq:environment tensor}) is in the form of a linear equation $(\mathbf{x}|\mathbf{A} = (\mathbf{b}|$, where $\mathbf{A}$ and $(\mathbf{b}|$ are known and $(\mathbf{x}|$ is an unknown vector of the environment matrices:
\begin{equation}
\begin{pmatrix}
(H_{i,0}| &
(H_{i,1}| &
(H_{i,2}|
\end{pmatrix}
\begin{pmatrix}
A_{0,0} & A_{0,1} & A_{0,2} \\
A_{1,0} & A_{1,1} & A_{1,2} \\
A_{2,0} & A_{2,1} & A_{2,2}
\end{pmatrix}
=
\begin{pmatrix}
(b_0| &
(b_1| &
(b_2|
\end{pmatrix},
\label{eq:GMRES}
\end{equation}
where the elements $A_{m,n}$ and $(b_n|$ can be read off from Eq. (\ref{eq:environment tensor}). In practice, we solve the set of equations defined by Eq. (\ref{eq:environment tensor}) for $m=0,1,2$ (or the equivalent Eq. (\ref{eq:GMRES})) using an iterative solver (for this work we use GMRES, but others may be employed).

\subsection{Fermions}
\label{subsec:fermions}

\begin{figure}
	\begin{subfigure}[c]{0.49\textwidth}
          \includegraphics[scale=0.25]{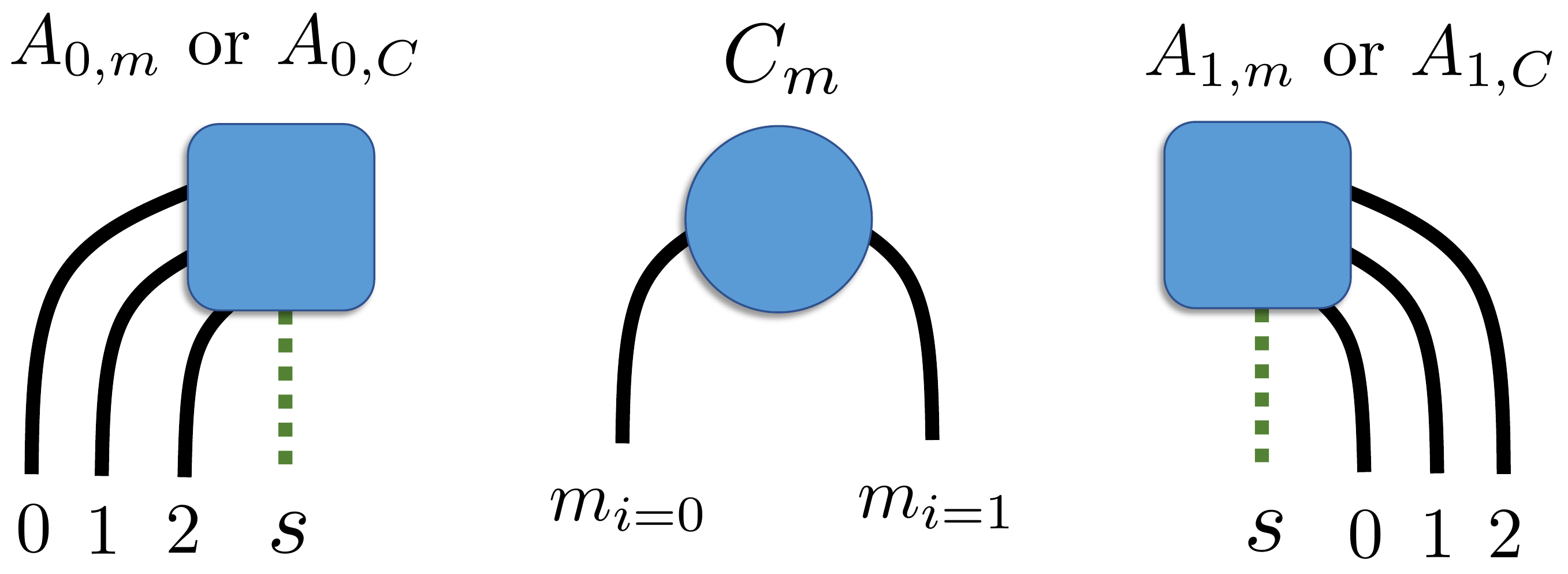}
          \caption{}
          \label{fig:fermionic ordering}
        \end{subfigure}\hspace{0.01\textwidth}%
	\begin{subfigure}[c]{0.49\textwidth}
          \includegraphics[scale=0.25]{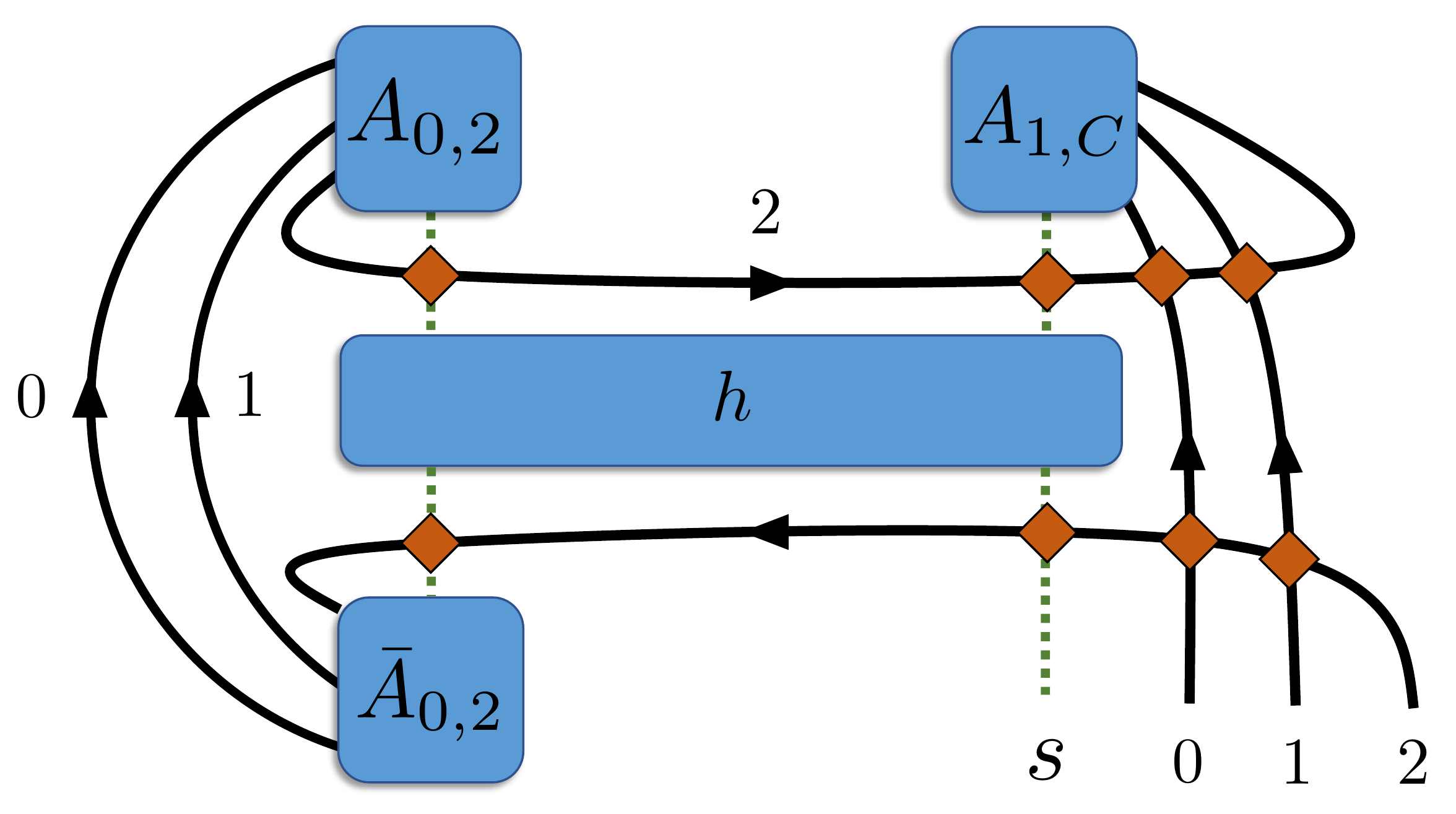}
          \caption{}
          \label{fig:fermionic ordering example}
        \end{subfigure}		
	\caption{(a) Fermionic ordering of the legs of the $A_{i,m}, A_{i,C}, C_m$ tensors. (b) An example of a tensor diagram containing fermionic swap gates. The fermionic swap gates are labeled by orange diamonds, and are present at the crossing of any two legs. Note that after the contraction of $A_{0,2}, \bar A_{0,2}, h$ and $A_{1,C}$, we have to put the dangling tensor legs back into the order of $A_{1,C}$, which requires three more swap gates.}
	\label{fig:fermionic ordering and example}
\end{figure}

As explained in the main text at the end of Section \ref{sec:VUTS}, for fermionic models like the Hubbard model, we need to use a fermionic version of VUTS. We use the method outlined in Refs. \cite{PhysRevB.81.165104,Orus_review}. Every tensor is endowed with a fermion parity $Z_2$ quantum number and is parity-preserving. When two tensor legs cross on a planar projection of a tensor diagram, a fermionic swap gate is placed at the crossing. In order to employ this method, we need to use a fixed ordering convention for the legs of the tensors $A_{i,m}, A_{i,C}, C_m$, which must be kept consistent in all of the diagrams in the calculation. The convention that we choose is shown in Fig. \ref{fig:fermionic ordering}. 
The legs of the tensors always appear in this order for tensor diagrams, regardless of their location in the diagram. An example of a diagram involving swap gates is shown in Fig. \ref{fig:fermionic ordering example}, where we draw a single term contributing to the action of the 1-site projected Hamiltonian on the 1-site wavefunction $A_{1,C}$ (the fermionic version of Fig. \ref{fig:H_Ac}). Please note that care has to be taken in how the diagrams are drawn, and also that the site equivalence of the system is properly obeyed. For more details on the method, such as additional diagrammatic rules that must be followed to account for the fermion signs, we refer the reader to Refs. \cite{PhysRevB.81.165104,Orus_review}.


\renewcommand\thefigure{B\arabic{figure}}
\renewcommand\thetable{B\arabic{table}}
\renewcommand\theequation{B\arabic{equation}}

\section{Details of the numerical calculation}
\label{sec:numerical details}

\subsection{Convergence strategies}
\label{sec:convergence strategies}

In this section we outline some of the numerical ingredients used in obtaining the phase diagram. For a generic value of $U$ and $\delta \mu$, we start the VUTS algorithm with a random ansatz of small bond dimension, $\chi = \chi_i \sim 4$. We then increase $\chi$ up to the desired $\chi_f$ using a generalization of the scheme explained in Appendix B of Ref. \cite{PhysRevB.97.045145}. In most of the phase diagram, any update schedule, i.e. any way of getting from $\chi_i$ to $\chi_f$, leads to VUTS converging to the same final state, within the precision error $\epsilon_{\text{prec}}$. We choose $\epsilon_{\text{prec}} \lesssim 10^{-6}$ for all our calculations. However, near the phase transition, $\delta \mu_c$, more care is needed in selecting the update schedule. Generically, we find that increasing $\chi$ in small increments and running several iterations after each small increase favors the insulating state, while using larger increments and a small number of iterations to get from $\chi_i$ to $\chi_f$ favors the metallic state. In this way, both branches near $\delta \mu_c$ can be obtained. Unfortunately, a given update schedule is never guaranteed to converge to a specific state, or to converge at all. Therefore, once the insulating/metallic state is found at a given $\delta \mu_0$, we can use that state as an ansatz for the state at $\delta \mu \approx \delta \mu_0$, in order to favor the phase of the $\delta \mu_0$ state. In this way, both branches can be reliably obtained for all $\delta \mu \approx \delta \mu_c$ (see Appendix \ref{sec:phase diagram more plots} for a discussion about continuing both branches into the regions of metastability). 
\\
\indent
The metallic ground state we find has vanishing magnetization, within our precision. In order to make sure we do not accidentally disfavor a magnetic metallic state by our choice of initial state, we do separate computations biased toward a magnetic metal by turning on a staggered external magnetic field, $H \rightarrow H - B_s (S_{(0)}^z - S_{(1)}^z)$, where the subscripts label the location in the unit cell. Then, we use this state as an ansatz for the metallic state close to $\delta \mu_c$ in the case when $B_s = 0$ (making sure that the density of the ansatz state is close to what we expect in the $B_s = 0$ case). Even with this method, we do not find any energetically favorable magnetic metal.

\subsection{Truncation error and run time}

\begin{figure}
	\begin{subfigure}[]{0.49\textwidth}
          \includegraphics[scale=0.9]{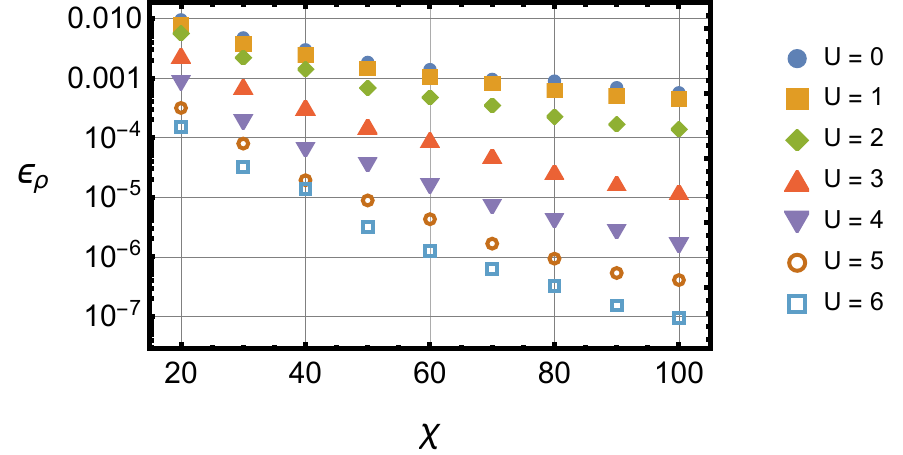}
          \caption{}
        \end{subfigure}\hspace{0.01\textwidth}%
	\begin{subfigure}[]{0.49\textwidth}
          \includegraphics[scale=0.8]{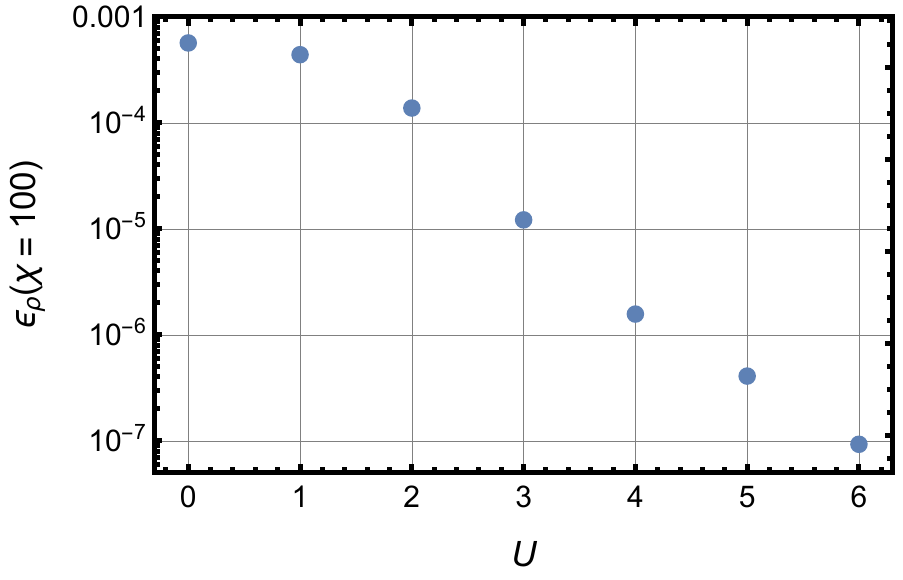}
          \caption{}
        \end{subfigure}
	\caption{Our estimate for the truncation error $\epsilon_\rho$ in the insulating phase (a) as a function of $\chi$ for several $U$ and (b) as a function of $U$ for the largest $\chi = 100$.}
	\label{fig:truncation error insulator}
\end{figure}

\begin{figure}
	\begin{subfigure}[]{0.49\textwidth}
          \includegraphics[scale=0.8]{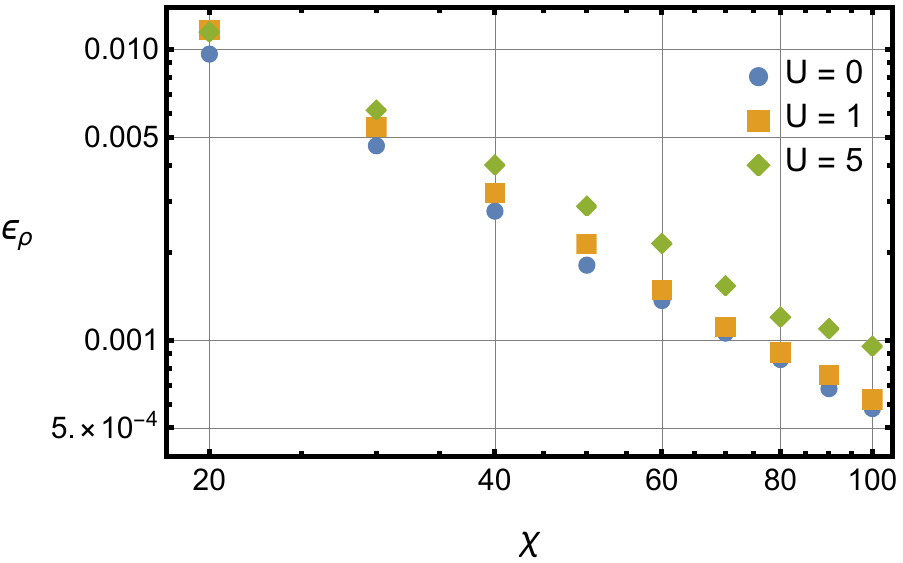}
          \caption{}
        \end{subfigure}\hspace{0.01\textwidth}%
	\begin{subfigure}[]{0.49\textwidth}
          \includegraphics[scale=0.8]{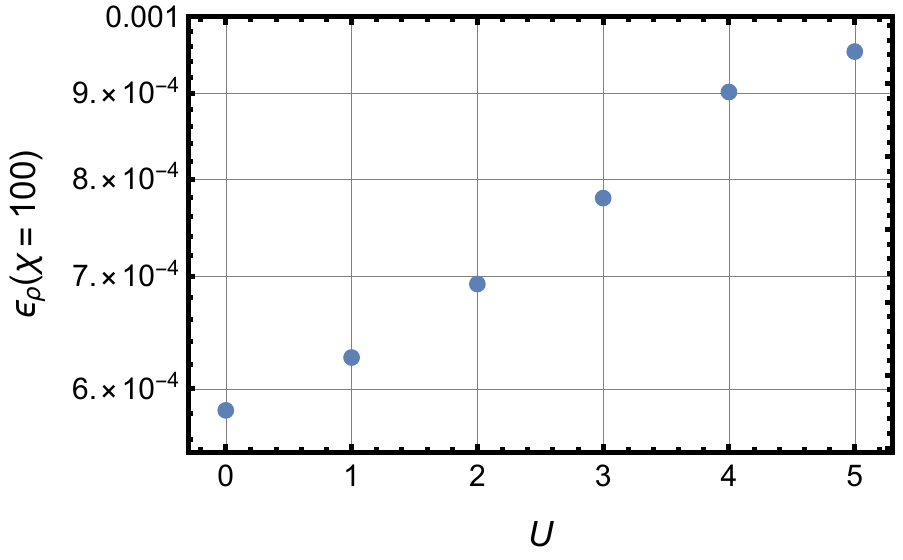}
          \caption{}
        \end{subfigure}
	\caption{Our estimate of the truncation error $\epsilon_\rho$ in the metallic phase at density $\<n_i(\delta \mu_c, \chi)\>$.}
	\label{fig:truncation error metal}
\end{figure}

Here we address the important question of numerical accuracy and run time. In tensor network states, the accuracy is measured by the truncation error or discarded weight $\epsilon_\rho$. In this work, we use a proxy for $\epsilon_\rho$, which is shown to be of the same order and used in Ref. \cite{PhysRevB.97.045145}, where it is referred to as $\norm{B_2}^2$ (we compute $\norm{B_2}^2$ using a trick similar to what is used in \cite{PhysRevB.86.195137}, in order for the computation to be manageable). Here we simply refer to $\norm{B_2}^2$ as $\epsilon_\rho$. In Fig. \ref{fig:truncation error insulator} we show the dependence of $\epsilon_\rho$ on $\chi$ and $U$ in the insulating phase ($\epsilon_\rho$ is essentially independent of $\delta \mu$ in the insulator).
We can see that (a) $\epsilon_\rho$ decays with $\chi$ roughly as an exponential for all $U$ and (b) $\epsilon_\rho$ decays with $U$ also roughly as an exponential. 
\\
\indent 
In the metallic phase, $\epsilon_\rho$ has a non-negligible dependence on the density, so for the purposes of comparison we look at densities $\<n_i\> = 1.2$ and the smallest density attainable for a given $U$ and $\chi$, $\<n_i(\delta \mu_c, \chi)\>$. In Fig. \ref{fig:truncation error metal} we plot $\epsilon_\rho$ for the smallest density as a function of $\chi$ for various $U$, and also as a function of $U$ at the largest $\chi = 100$ (the corresponding plots for $\<n_i\> = 1.2$ are in Fig. \ref{fig:truncation error metal n = 1.2}).
We can see that, as in Fig. \ref{fig:truncation error metal n = 1.2}, the decay with $\chi$ is algebraic, as expected for a gapless system. As a function of $U$, $\epsilon_\rho$ increases initially and then seems to saturate, although the large $U$ behavior is undetermined. 
\\
\indent
To conclude, we note that for the largest bond dimension we present, $\chi = 100$, $\epsilon_\rho < 10^{-3}$ for every point in parameter space we study.  
\\
\indent 
In order to improve the run time, we use multithreading over the block sparse tensors, which speeds up the calculation significantly. The largest bond dimension we present in this work is $\chi = 100$. We thread over $12,24$ and $48$ cores, for which the run times for a single VUTS iteration at $\chi = 100$ are $24.2,18.5$ and $15$ minutes, respectively. The number of iterations required to achieve a precision error of $\epsilon_{\text{prec}} \lesssim 10^{-6}$ depends significantly on the proximity of the ansatz wavefunction to the true ground state, and could be as low as tens of iterations and as large as more than a thousand iterations. Therefore, it is very important to have good convergence strategies, as described in the previous section.


\renewcommand\thefigure{C\arabic{figure}}
\renewcommand\thetable{C\arabic{table}}
\renewcommand\theequation{C\arabic{equation}}

\section{Supporting plots for phase diagram and discussion of first-order transition}
\label{sec:phase diagram more plots}

\begin{figure}
	\begin{subfigure}[]{0.49\textwidth}
          \includegraphics[scale=0.9]{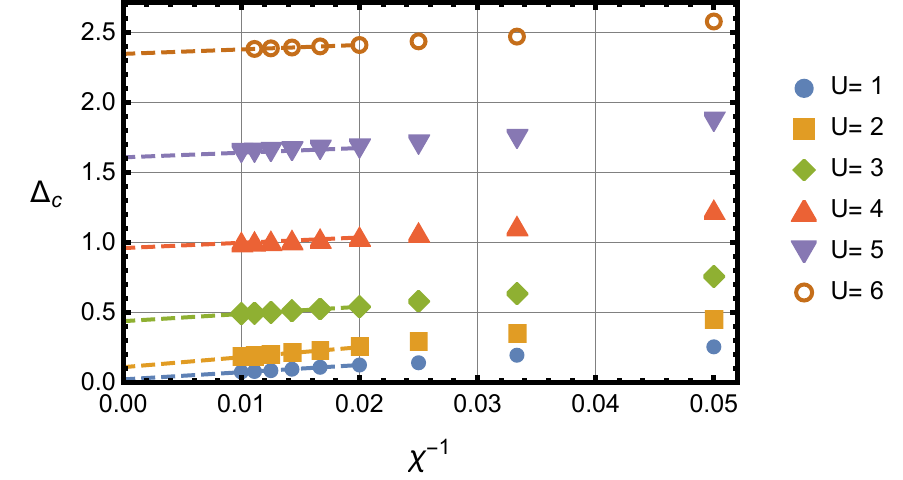}
	      \caption{The charge gap $\Delta_c (\chi)$ as a function of increasing $\chi$.}
	      \label{fig:charge gap finite chi}
        \end{subfigure}\hspace{0.01\textwidth}%
	\begin{subfigure}[]{0.49\textwidth}
          \includegraphics[scale=0.9]{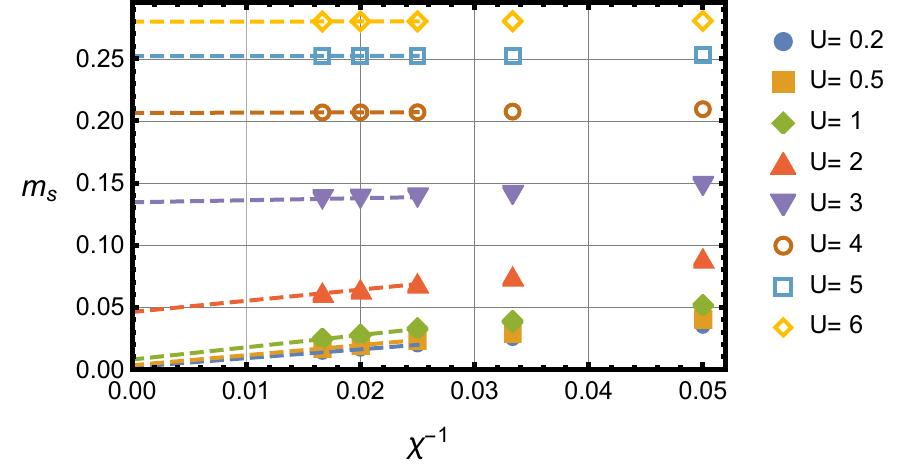}
	      \caption{The staggered magnetization $m_s (\chi)$ of the insulating state at half filling as a function of increasing $\chi$.}
          \label{fig:ms finite chi}
        \end{subfigure}\hspace{0.01\textwidth}%
	\begin{subfigure}[]{0.49\textwidth}
          \includegraphics[scale=0.9]{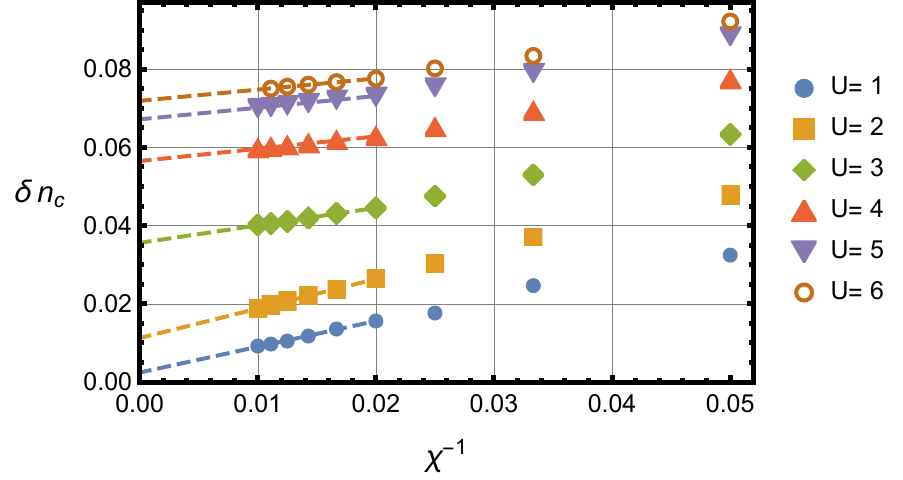}
	      \caption{The jump in occupation at the transition point, $\delta n(\delta \mu_c, \chi) $ as a function of increasing $\chi$.}
	      \label{fig:jump in occupation finite chi}
        \end{subfigure}\hspace{0.01\textwidth}%
	\begin{subfigure}[]{0.49\textwidth}
          \includegraphics[scale=0.85]{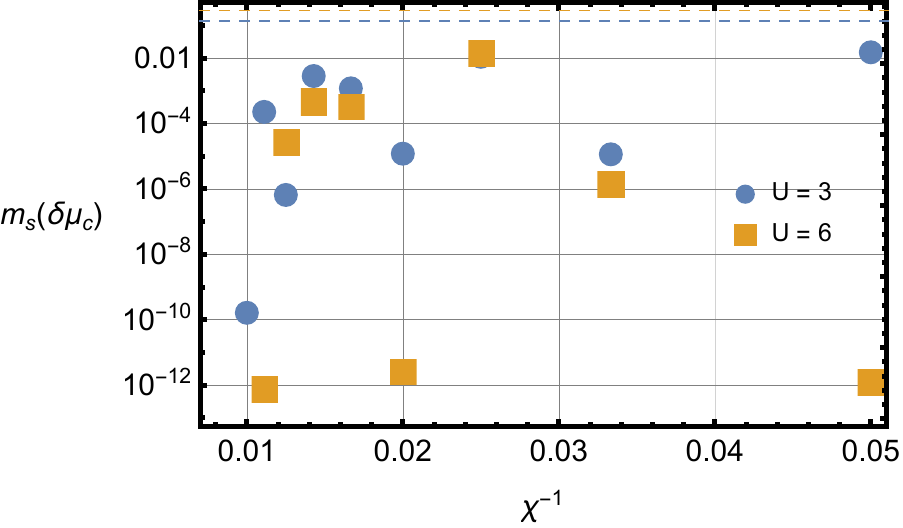}
	      \caption{The staggered magnetization $m_s$ of the metallic state at the critical point versus the inverse bond dimension for $U = 3,6$. The horizontal dashed lines indicate the magnetization of the insulating state across the phase transition, computed at $\chi = 100$.}
	      \label{fig:ms at QP vs chi}
        \end{subfigure}\hspace{0.01\textwidth}%
    \caption{Several physical quantities plotted as functions of $\chi$. Plots (a),(b),(c) include linear fits.}
	\label{fig:several plots finite chi}
\end{figure}

\begin{figure}
	\includegraphics[scale=0.7]{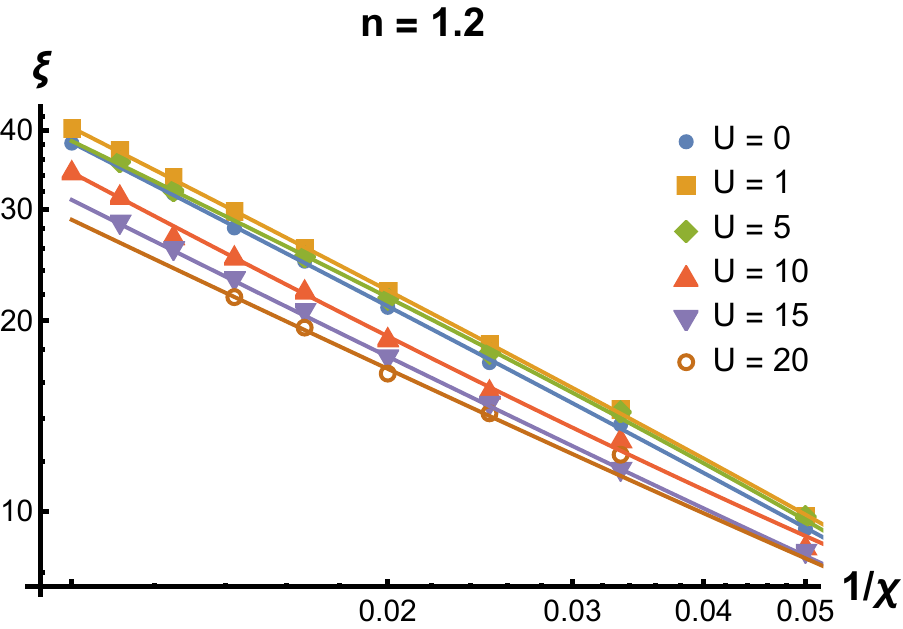}
	\caption{The correlation length scaling with $\chi$ for a range of $U$ and $\<n_i\> = 1.2$. The fits shown are of the form $\xi \sim \chi^{\a}$ with $0.7 < \a < 1$.}
	\label{fig:xi vs chi all U at n = 1.2}
\end{figure}

\begin{figure}
    \begin{subfigure}[c]{0.49\textwidth}
      \includegraphics[scale=0.85]{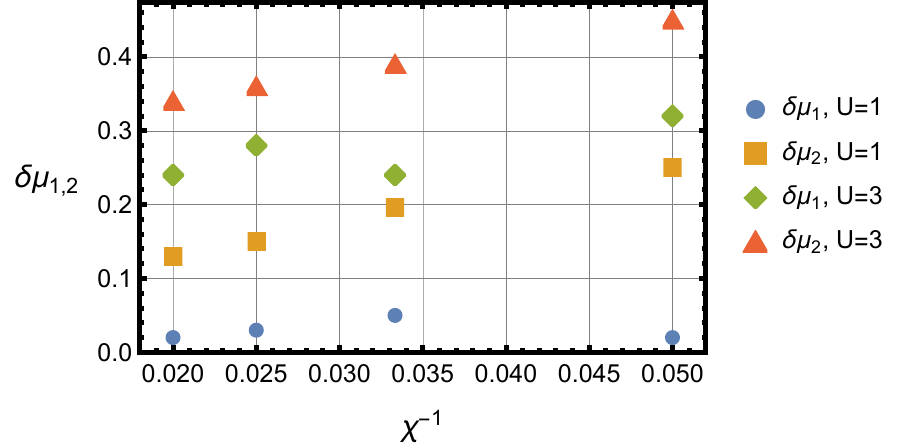}
      \caption{}
    \end{subfigure}\hspace{0.01\textwidth}%
    \begin{subfigure}[c]{0.49\textwidth}
      \includegraphics[scale=0.85]{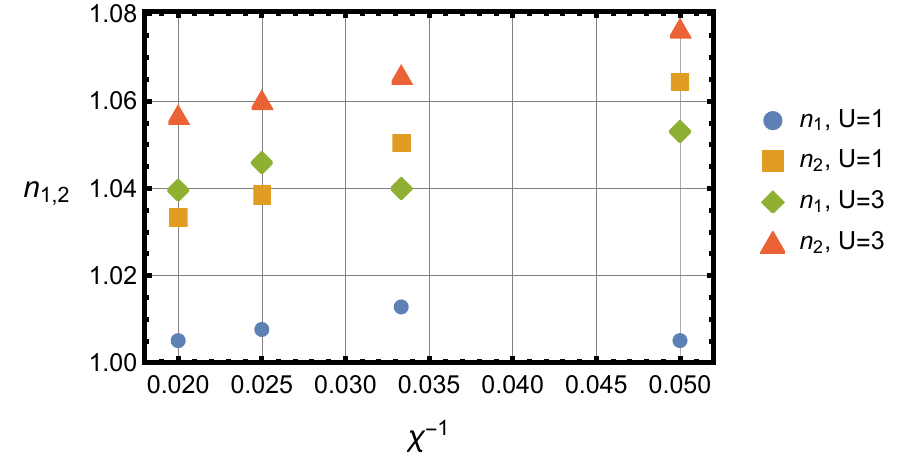}
      \caption{}
    \end{subfigure}
	\caption{The spidonal values $\delta \mu_1, \delta \mu_2$ and the spinodal values of density $n_1, n_2$ for $U = 1, 3$.}
	\label{fig:delta mu_i and n_i}
\end{figure}

Here we provide more supporting plots of finite $\chi$ results from Sec. \ref{sec: Hubbard model}, and also discuss in more detail the first-order metal-insulator phase transition, including the calculation of the charge gap. 
\\
\indent 
We start by illustrating the behavior of the density as a function of $\delta \mu$. We show this in Fig. \ref{fig:occ vs mu U = 2,5} for $U = 2,5$ and a range of $\chi$. All other values of $U$ and $\chi$ behave qualitatively similar. 
\begin{figure}
	\begin{subfigure}[]{0.49\textwidth}
          \includegraphics[scale=0.9]{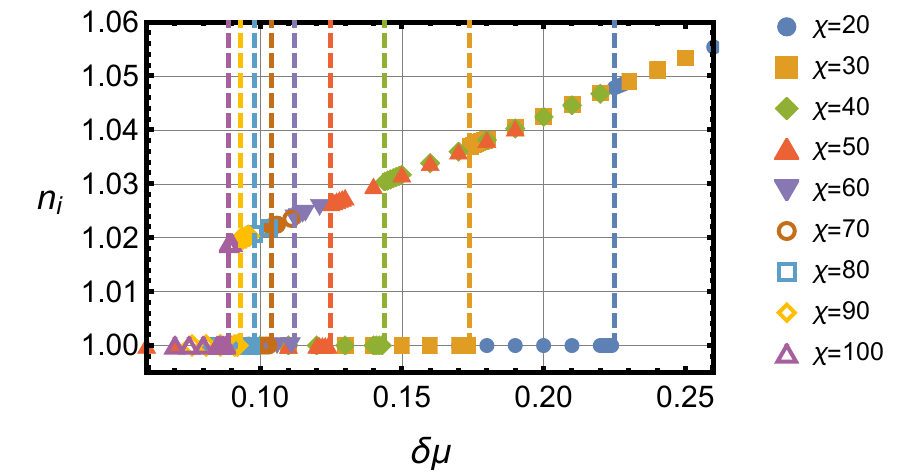}
          \caption{$U = 2$.}
        \end{subfigure}\hspace{0.01\textwidth}%
	\begin{subfigure}[]{0.49\textwidth}
          \includegraphics[scale=0.9]{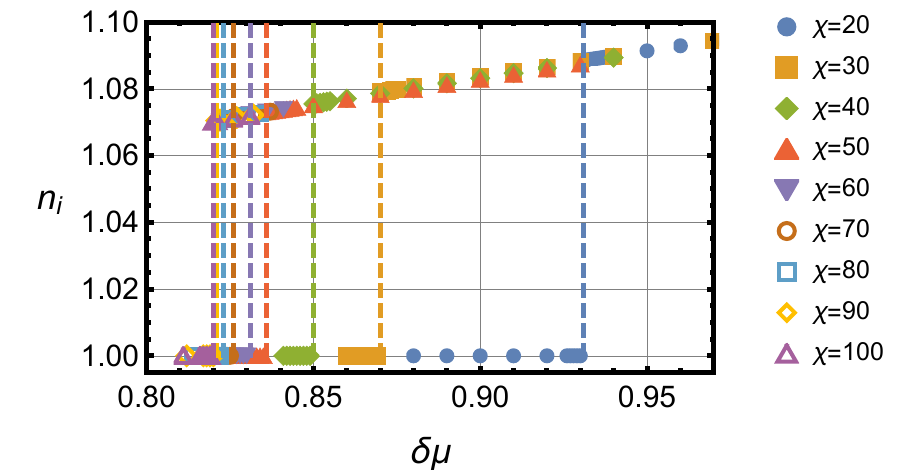}
          \caption{$U = 5$.}
        \end{subfigure}
	\caption{Fermion occupation $\< n_{i} \>$ as a function of $\delta \mu$ for $U = 2,5$ and a range of $\chi$.}
	\label{fig:occ vs mu U = 2,5}
\end{figure}
Now, in Fig. \ref{fig:several plots finite chi} we show several finite $\chi$ plots, the extrapolations of which are in the main text.
In Figs. \ref{fig:charge gap finite chi}, \ref{fig:ms finite chi}, and \ref{fig:jump in occupation finite chi} we plot the charge gap, staggered magnetization at half filling, and the jump in occupation at the phase transition, respectively, as a function of $\chi$ for a range of $U$. We fit the data with a linear fit, ignoring the first few points, after which the plots data look linear. The extrapolated values are shown in Figs. \ref{fig:charge gap extrapolation}, \ref{fig:ms extrapolation}, and \ref{fig:jump in occupation extrapolation}, respectively. In Fig. \ref{fig:ms at QP vs chi} we plot the finite $\chi$ values of the staggered magnetization on the metallic side of the phase transition for some values of $U$ (other values of $U$ behave similarly). We can see the (non-monotonic) trend of $m_s(\chi)$ decreasing to arbitrarily small values as $\chi$ increases. For completeness we also provide the scaling of the correlation length $\xi(\chi)$ with $\chi$ for a range of $U$ at density $\<n_i\> = 1.2$, shown in Fig. \ref{fig:xi vs chi all U at n = 1.2}.
\\
\indent 
Now we turn to the question of the first order metal-insulator transition. Near the transition point $\delta \mu_c$, both the insulating and metallic branches exist and can be stabilized for a range of $\delta \mu$, as can be seen in Fig. \ref{fig:n and energy vs delmu for the two branches at U = 6 and chi = 50}. The meta-stable parts of the branches are obtained by way of ``hysteresis" in the numerical algorithm. We start, e.g. with the insulating ground state at a given $\delta \mu_x < \delta \mu_c$ and use that as an ansatz for $\delta \mu_y > \delta \mu_c$, then use the insulating solution thus found at $\delta \mu_y$ as an ansatz for $\delta \mu_z > \delta \mu_y$ and so on. The metallic branch is obtained in reverse. This approach is appealing because of its similarity to the way actual hysteresis curves in a laboratory are obtained (another approach that can be parallelized is to use bond dimension growth schedules that favor one branch or the other (c.f. Sec. \ref{sec:convergence strategies})). However, this method of obtaining the spinodals $\delta \mu_1, \delta \mu_2$ (see Section \ref{sec: Hubbard model} for definition) will very likely overshoot $\delta \mu_2$ and undershoot $\delta \mu_1$. In other words, we are not at all guaranteed to obtain both branches for the true width of the hysteresis window. Nonetheless, in Fig. \ref{fig:delta mu_i and n_i} we show how our computed $\delta \mu_1, \delta \mu_2$ behave as a function of $\chi$ for some values of $U$, as well as the corresponding spinodal values of density, $n_1 = n(\delta \mu_1), n_2 = n(\delta \mu_2)$. 
We can see that, although there is a general trend, $\delta \mu_i$ and $n_i$ are not always smooth as functions of $\chi$. Even though we cannot find the true spinodal values, the mere existence of a hysteresis of this type is another indication of the first-order nature of the transition \cite{Yan1173}. Furthermore, Fig. \ref{fig:delta mu_i and n_i} suggests that in the $\chi \to \infty$ limit the left spinodal value $n_2$ might vanish for small values of $U$ (although, as we stated, since the behavior is not monotonic we are not doing any extrapolation). If there is a way to stabilize this branch (perhaps with finite temperature or some other parameter), this would imply that the transition could become continuous below a certain value of $U$, which would be very interesting. 
\\
\indent 
Before ending the discussion of the meta-stable states, we mention another point. As can be seen from Fig. \ref{fig:occ vs mu U = 2,5}, the behavior of the density is quite linear, particularly near $\delta \mu_c$. If we extrapolate this linear behavior, we can define a value $\delta \mu^*$ at which the extrapolation intersects the $\delta \mu$ axis. $\delta \mu^*$ is the lowest possible value of $\delta \mu _2$. Interestingly, as a function of $U$ the value of $\delta \mu^*$ changes from zero to non-zero around a certain value $U^*$. The extrapolation in $\chi$ is difficult to do, due to non-monotonicity, but we can estimate that $3 < U^* < 4$. This is reminiscent of the behavior found from DMFT in the $z=\infty$ limit, where the metallic spinodal terminates at $\delta\mu_2=0$ for $U<U_{c2}$ and at a finite value of $\delta\mu_2$ for $U>U_{c2}$ ($U_{c2}$ being a finite critical value) \cite{RevModPhys.68.13}.
\\
\indent 
Finally, we turn to the question of computing the charge gap. The usual formula is $\Delta_c = E_0(N+1) + E_0(N-1) - 2 E_0(N)$, where $E_0(N)$ is the canonical ground state energy with $N$ particles, and $E_0(N\pm1)$ are the canonical energies of the metallic ground states with $N \pm 1$ particles, respectively. In this case, it is not immediately clear what is the energy $E_0(N\pm1)$, since there is no metallic state with density arbitrarily close to one in the grand canonical ensemble (in which our calculation is performed). However, in the canonical ensemble, we can take advantage of phase separation, and insert or remove a particle by creating a mixture of the two phases in the system. Indeed, working at finite volume $V$, at the critical point $\delta \mu_c$ we have an insulator with density $n_I = 1$ and a metal with density $n_M = 1 \pm \delta n$ (the $\pm$ for adding and removing a particle). The fraction of the system in the insulating phase is given by $\alpha = 1 - \frac{1}{\delta n \, V}$ (the same for adding or removing a particle, because of particle-hole symmetry). Computing the charge gap then gives $\Delta_c = \frac{2(\eps_M - \eps_I)}{\delta n}$, where $\eps_M, \eps_I$ are the canonical energy densities of the metallic and insulating states, respectively. The transition point $\delta \mu_c$ is defined such that the grand canonical energy densities of the two states are equal, which gives $\eps_M - \delta \mu_c (n_I + \delta n) = \eps_I - \delta \mu_c \, n_I$. This gives $\Delta_c = 2 \delta \mu_c$, which is the naive expectation.


\renewcommand\thefigure{D\arabic{figure}}
\renewcommand\thetable{D\arabic{table}}
\renewcommand\theequation{D\arabic{equation}}

\section{Occupation in $\theta$-space}
\label{sec:occupation function}

Here we derive Eq. (\ref{eq:n_theta final}). From Eqs. (\ref{eq:occupation function definition}), (\ref{eq:theta transform c operators}) we have
\begin{equation}
n(\theta, \theta') = \lim\limits_{L \to \infty}\frac{\pi}{L+1} \sum_{d,d' = 0}^{L} \psi_d(\th) \psi_{d'}(\th') \, \< \td c^{\dag}_{d'} \td c_{d} \>. 
\label{eq:n_theta real space insertion 1}
\end{equation}
We need to remember that the operator $\td c_{d}$ implies a reference to a center site, which we always keep the same. Now we expand $\< \td c^{\dag}_{d'} \td c_{d} \>$ in operators $c_i$, where $i$ labels a site on the original Bethe lattice. We show an example in Fig. \ref{fig:BL 4 generations}. We use the fact that $\< c^{\dag}_{i} c_{j} \> = \< c^{\dag}_{j} c_{i} \>$ is a function of $|i-j|$, which is the length of the path connecting sites $i,j$. We first consider the case when $d,d' > 0, d \neq d'$. Since $d, d'$ label generations emanating from the same center site, the correlation function $\< \td c^{\dag}_{d'} \td c_{d} \>$ becomes
\begin{equation}
\< \td c^{\dag}_{d'} \td c_{d} \> = \frac{z (z-1)^{\min(d,d')-1}}{z \sqrt{(z-1)^{d+d'-2}}} \sum_{r=0}^{\min(d,d')} A_r \, \< c^{\dag}_{0} c_{\abs{d-d'}+2r} \>.
\label{eq:c^dag_d c_d'}
\end{equation}
The denominator outside of the sum comes from the normalization factor of the symmetric states. The numerator outside of the sum is the number of sites of the inner shell. The coefficients $A_r$ are the number of sites on the outer shell that are connected to a given site on the inner shell by a distance $\abs{d-d'}+2r$. To find $A_r$ we start with $r = 0$. The number of these is $A_{0} = (z-1)^{\abs{d-d'}}$. Now we increase $r$ by one, which means we go from the given site on the inner shell up one generation towards the center site, and then back down a different branch. Going back down gives us a factor of $z-2$ to get back to the inner shell, and then another factor of $(z-1)^{\abs{d-d'}}$. For $r = 2$, we need to go up two generations towards the center. Coming back down now gives us $(z-2) \, (z-1) \, (z-1)^{\abs{d-d'}}$. For a general $0 < r < \min(d,d')$ this gives $A_r = (z-2) \, (z-1)^{r-1} \, (z-1)^{\abs{d-d'}}$. The last term in the series is special since moving up one generation bring us to the center site, so the factor $z-2$ becomes $z - 1$ and $A_{\min(d,d')} = (z-1)^{\min(d,d')} \, (z-1)^{\abs{d-d'}}$. We can rewrite Eq. (\ref{eq:c^dag_d c_d'}) as 
\begin{align}
\nn \< \td c^{\dag}_{d'} \td c_{d} \>  & 
=
\sqrt{(z-1)^{\abs{d-d'}}} \,  
\sum_{r=0}^{\min(d,d')} \(\frac{z-1}{z-2}\)^{\delta_{r,0} + \delta_{r,\min(d,d')}} (z-2)(z-1)^{r-1} \, \< c^{\dag}_{0} c_{\abs{d-d'}+2r} \>
\\ & = 
\frac{z-2}{\sqrt{z (z-1)}} \,
\sum_{r=0}^{\min(d,d')} \(\frac{z-1}{z-2}\)^{\delta_{r,0} + \delta_{r,\min(d,d')}} \, \< \td c^{\dag}_{0} \td c_{\abs{d-d'}+2r} \>.
\label{eq:c^dag_d c_d' simplified}
\end{align}
Now we look at the special cases. If $d = d' \neq 0$, everything goes through the same except the conversion from $\< c^{\dag}_{0} c_{\abs{d-d'}+2r} \>$ to $\< \td c^{\dag}_{0} \td c_{\abs{d-d'}+2r} \>$, which gives
\begin{equation}
\nn \< \td c^{\dag}_{d} \td c_{d} \> 
= 
\frac{z-2}{\sqrt{z (z-1)}} \, \sum_{r=0}^{d} \(\frac{z-1}{z-2}\)^{\delta_{r,0} + \delta_{r,d}} \(\frac{z}{z-1}\)^{\frac{\delta_{r,0}}{2}} \, \< \td c^{\dag}_{0} \td c_{2r} \>.
\label{eq:c^dag_d c_d}
\end{equation}
If $d' = 0, d \geq 0$ (equivalent to $d' \geq 0, d = 0$), the relation should just trivially give back $\< \td c^{\dag}_{0} \td c_{d}\>$. Incorporating these two special cases, Eq. (\ref{eq:c^dag_d c_d' simplified}) can be written as
\begin{equation}
\< \td c^{\dag}_{d'} \td c_{d} \> 
= 
\frac{z-2}{\sqrt{z (z-1)}} \,
\(\frac{\sqrt{z}(z-2)}{(z-1)^{3/2}}\)^{\delta_{\min(d,d'),0}} 
\sum_{r=0}^{\min(d,d')} \(\frac{z-1}{z-2}\)^{\delta_{r,0} + \delta_{r,\min(d,d')}} 
\, \sqrt{\frac{z}{z-1}}^{\delta_{d,d'} \delta_{r,0} - \delta_{d,0} \delta_{d',0}} 
\, \< \td c^{\dag}_{0} \td c_{\abs{d-d'}+2r} \>.
\label{eq:c^dag_d c_d' 2}
\end{equation}
Important to note, this is not a function of $\abs{d-d'}$ only. Therefore, we cannot simplify this further by doing the summation over $d+d'$. The final expression is given by inserting Eq. (\ref{eq:c^dag_d c_d' 2}) into Eq. (\ref{eq:n_theta real space insertion 1}) and shown in Eq. (\ref{eq:n_theta final}).

\end{document}